\documentclass[prl,twocolumn,showpacs,amsmath,amssymb,superscriptaddress]{revtex4-1}
\usepackage{graphicx}
\usepackage{color}
\usepackage{braket}
\usepackage{amsmath}
\usepackage{amsfonts}

\usepackage{amsmath,amssymb,mathrsfs}
\usepackage{bbm}
\usepackage[dvipsnames]{xcolor}

\date{\today}                  

\begin{document}

\title{Universality of Boundary Charge Fluctuations}

\author{Clara S. Weber}
\affiliation{Institut f\"ur Theorie der Statistischen Physik, RWTH Aachen, 
52056 Aachen, Germany and JARA - Fundamentals of Future Information Technology}

\author{Kiryl Piasotski}
\affiliation{Institut f\"ur Theorie der Statistischen Physik, RWTH Aachen, 
52056 Aachen, Germany and JARA - Fundamentals of Future Information Technology}

\author{Mikhail Pletyukhov}
\affiliation{Institut f\"ur Theorie der Statistischen Physik, RWTH Aachen, 
52056 Aachen, Germany and JARA - Fundamentals of Future Information Technology}

\author{Jelena Klinovaja}
\affiliation{Department of Physics, University of Basel, Klingelbergstrasse 82, 
CH-4056 Basel, Switzerland}
\author{Daniel Loss}
\affiliation{Department of Physics, University of Basel, Klingelbergstrasse 82, 
CH-4056 Basel, Switzerland}
\author{Herbert Schoeller}
\affiliation{Institut f\"ur Theorie der Statistischen Physik, RWTH Aachen, 
52056 Aachen, Germany and JARA - Fundamentals of Future Information Technology}
\author{Dante M. Kennes}
\email[Email: ]{Dante.Kennes@rwth-aachen.de}
\affiliation{Institut f\"ur Theorie der Statistischen Physik, RWTH Aachen, 
52056 Aachen, Germany and JARA - Fundamentals of Future Information Technology}
\affiliation{Max Planck Institute for the Structure and Dynamics of Matter, Center for Free Electron Laser Science, 22761 Hamburg, Germany}

\begin{abstract}
We establish the quantum fluctuations $\Delta Q_B^2$ of the charge $Q_B$ accumulated at the boundary of an insulator as an integral tool to characterize phase transitions where a direct gap closes (and reopens), typically occurring for insulators with topological properties. The power of this characterization lies in its capability to treat different kinds of insulators on equal footing; being applicable to transitions between topological and non-topological band, Anderson, and Mott insulators alike. In the vicinity of the phase transition we find a universal scaling $\Delta Q_B^2(E_g)$ as function of the gap size $E_g$ and determine its generic form in various dimensions. 
For prototypical phase transitions with a massive Dirac-like bulk spectrum we demonstrate a scaling with the inverse gap in one dimension and a logarithmic one in two dimensions.   
\end{abstract}


\maketitle



\emph{Introduction ---}
In the last few decades studies concerning topological phases of matter, i.e.~phases not characterized by a Landau-type of order parameter, have moved to the vanguard of condensed matter research \cite{volkov_pankratov_jetp_85,pankratov_etal_SSC_87,kane_mele_prl_95,bernevig_etal_science_06,fu_kane_mele_prl_07,koenig_etal_science_07,hsieh_etal_nature_08,hasan_kane_RMP_10,qi_zhang_RMP_11,bernevig_book_13,tkachov_book_15,asboth_etal_book_16}. A topological phase transition separates two insulating phases with different topological properties and is typically accompanied by a band inversion at a special point in quasimomentum space where two bands are directly coupled. Whereas standard metal-insulator transitions are described via localization theories \cite{kohn_pr_64,resta_jpc_02}, a topological phase transition probes specific low-energy features and is characterized by a closing and reopening of a direct gap, accompanied by a change of a topological index. 
Independent of whether a topological index remains the same or not at such a transition, the fundamental question arises how to embed these special phase transition into conventional ones, where the fluctuations of an appropriate observable diverge at the transition, accompanied by the divergence of a characteristic length scale. Close to the transition such a diverging length scale is naturally given by $\xi_g=v_F/E_g$, where $v_F$ is a typical velocity and $E_g$ denotes the gap size.  Going one step further this poses the interesting issue whether fluctuations reveal universal scaling laws as function of $\xi_g$ (or, equivalently, $E_g$).

\begin{figure}[t!]
    \centering
	 \includegraphics[width =0.9\columnwidth]{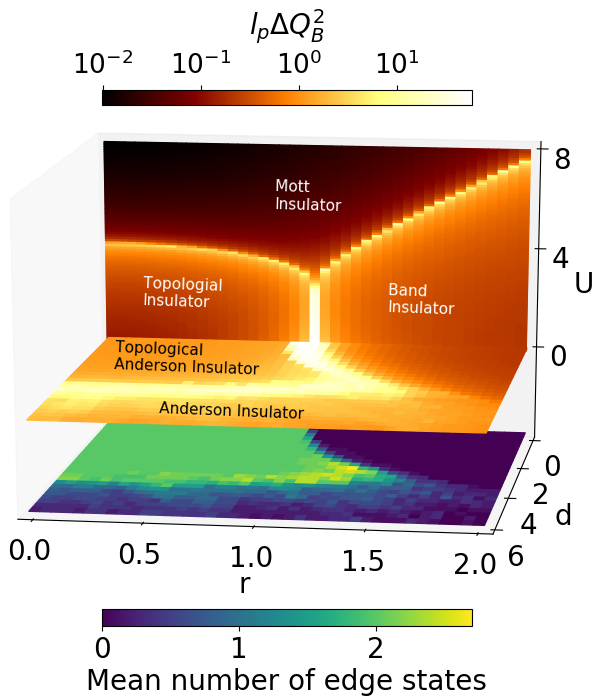}
	  \caption{Topological phase diagram characterized by the fluctuations $l_p \Delta Q_B^2$ (top) and by the number of zero-energy edge states (bottom) for the SSH model studied experimentally in \cite{meier_etal_science_18}. 
	  $d$ is disorder strength and $U$ denotes nearest-neighbor Coulomb interaction. $r=t_1/t_2<1$ defines the topological region for $d=U=0$. Phase boundaries between topological and non-topological band, Anderson, and Mott insulators are all well characterized by strongly enhanced fluctuations \cite{Fig1_para}. 
	  }
    \label{fig:phase_diagram}
\end{figure}

Recently it has been proposed that the boundary charge $Q_B$ accumulated at a $D-1$-dimensional flat surface of a $D$-dimensional insulator probes universal properties of topological insulators at low energies  \cite{park_etal_prb_16,thakurathi_etal_prb_18,pletyukhov_etal_prbr_20,pletyukhov_etal_prb_20,pletyukhov_etal_prr_20,lin_etal_prb_20}.
Close to the transition point, it was demonstrated for one-dimensional, single-channel models that $Q_B$ directly probes the phase of the gap parameter (in units of $2\pi$) independent of the gap size, and reveals half-integer jumps at  Weyl semimetal-like transitions \cite{pletyukhov_etal_prbr_20,pletyukhov_etal_prb_20,pletyukhov_etal_prr_20}. Therefore, one expects strong fluctuations of $Q_B$ at a (topological) phase transition and it is quite surprising that these fluctuations have so far not drawn much attention \cite{park_etal_prb_16}. 

We remedy this substantial oversight in this letter and demonstrate that the fluctuations $\Delta Q_B^2=\langle \hat{Q}_B^2\rangle - \langle \hat{Q}_B\rangle^2$ of the boundary charge themselves are the key to addressing universal properties of (topological) phase transitions. We identify a {\it universal regime} $l_p\gg\xi_g\gg a$ where $l_p\Delta Q_B^2(\xi_g)$ is a universal function of $\xi_g$, i.e., independent of the microscopic details of the charge measurement probe, described by a macroscopic length scale $l_p$ on which the probe looses the contact to the sample (see below). Universality implies independence from high-energy properties, relevant on the scale of the lattice spacing $a$. In the regime close to the phase transition ($\xi_g\to\infty$) we find that $l_p\Delta Q_B^2$ diverges in one and two dimensions, quite analog to divergent fluctuations in conventional phase transitions. Therefore, we suggest the fluctuations of $Q_B$ as a useful and measurable tool to probe the phase diagram of topological insulators. In Fig.~\ref{fig:phase_diagram} we begin by a compelling demonstration of the power of the suggested characterization focusing on the prototypical Su-Schrieffer-Heeger (SSH)  model \cite{su_etal_prl_79,su_etal_prb_80} at half-filling including hopping disorder (experimentally studied in Ref.~\cite{meier_etal_science_18}) and nearest neighbor Coulomb interaction. Details of the model are postponed to Eq.~\eqref{eq:model}, however, the general physics is dictated by the topological index $r$ being the ratio of the two hopping amplitudes in the SSH model. Without disorder and interaction, the topologically non-trivial phase transitions to a trivial one at $r=1$. Including disorder in the hoppings $d$ a topological Anderson insulator is stabilized even beyond $r=1$  for weak disorder, while for strong disorder a trivial Anderson insulator is found. In the presence of strong enough Coulomb interaction $U$ a Mott insulator is established {\color{black} (for an interpretation of the Mott transition as a topological one see Ref.~\cite{sen_etal_prb_20})}. All of the different phase boundaries between topological and non-topological band, Anderson, and Mott insulators are signalled by diverging boundary charge fluctuations $l_p \Delta Q_B^2\sim\xi_g$. Our approach thus unifies transitions between all of these different classes of single-particle and correlated insulators. 
In addition, below, we find that $l_p \Delta Q_B^2$ shows a  universal scaling as function of $E_g$ for a variety of models. We find a striking dependence on the dimensionality of the system which we exemplify for a massive Dirac-like low-energy spectrum. We report a typical scaling with the inverse gap in one dimension, logarithmic scaling in two-dimensional systems, and a monotonic increase of the fluctuations to a finite value at zero-gap in three dimensions.


\begin{figure}[t]
    \centering
	 \includegraphics[width =\columnwidth]{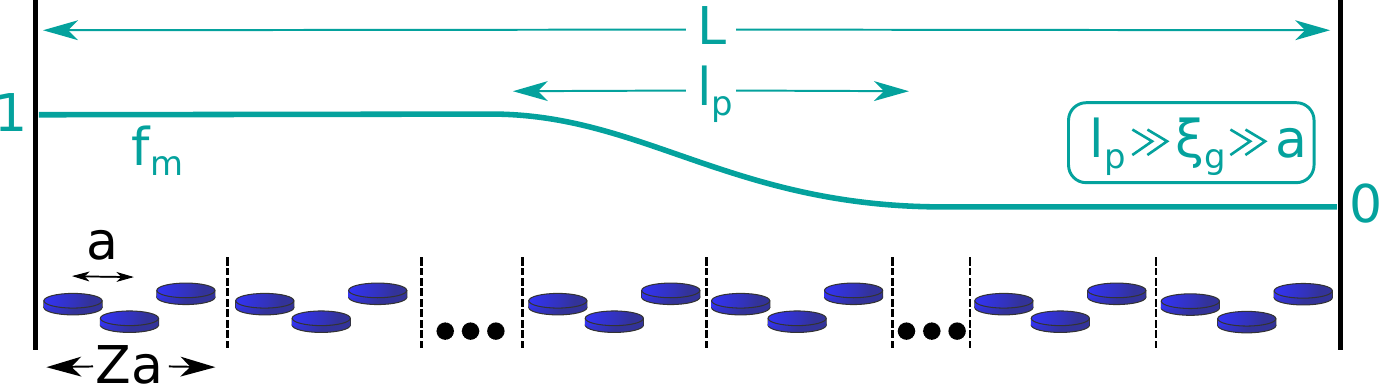}
	  \caption{Pictorial representation of the system (bottom) and the definition of the envelope of the charge probe $f_m$ (top). {\color{black} Here, $l_p$ defines the length on which the probe smoothly looses contact to the sample (region where envelope varies smoothly from $1$ to $0$). We also define the lattice spacing $a$ and the unit cell size $Za$. Implicitly, we assume that the fall-off of $f_m$ fits into the system size $L$ (the center of the fall-off is irrelevant). The universal regime is defined by $l_p\gg\xi_g=v_F/E_g\gg a$.  }
	  }
    \label{fig:sys}
\end{figure}

\emph{Model and  Boundary charge fluctuations---}
We consider a generic and finite $D$ dimensional tight-binding model with a $D-1$ dimensional flat surface. In the direction perpendicular to the surface we consider $N_s=L/a$ lattice sites labeled by $m=1,\dots,N_s$ with open boundary conditions, where $L$ denotes the system size. We take an arbitrary extend and periodic boundary conditions in the remaining (transverse) directions. We then define an effective one-dimensional chain with two ends by absorbing the remaining directional degrees of freedom into a (large) multi-channel character on each site defining $N_B$ transverse channels labeled by $\sigma=1,\dots,N_B$ (which can additionally include, e.g., spin or orbital degrees of freedom as well). The size of the unit cell of the effective one-dimensional chain is denoted by $Za$, and $x=ma$ defines the position of lattice site $m$; see Fig.~\ref{fig:sys} (bottom). We consider zero temperature and fixed particle number $N$, and concentrate on the low-energy limit where the gap $E_g$ is assumed to be small compared to the bandwidth or, equivalently, $\xi_g=v_F/E_g\gg a$. Generalizations are discussed in the SM \cite{SM}. We choose units $\hbar=e=1$.

The boundary charge is a macroscopic observable measured on scales much larger than the microscopic scale $\xi_g\gg a$. We describe
the macroscopic average by an envelope function $f_m$ characteristic for a charge measurement probe, which falls off smoothly from unity to zero on the macroscopic length scale $l_p\gg\xi_g$, see Fig.~\ref{fig:sys} (top). The boundary charge operator at one end {\color{black}(determined by the envelope function $f_m$ falling off from that end)} of the system is defined~\cite{park_etal_prb_16,pletyukhov_etal_prb_20} by $\hat{Q}_B = \sum_{m=1}^{N_s} f_m (\hat{\rho}_m - N/N_s)$, where $\hat{\rho}_m=\sum_\sigma a^\dagger_{m\sigma}a_{m\sigma}$ is the charge operator at site $m$ summed over all $N_B$ channels. The fluctuations $\Delta Q_B^2=\langle\hat{Q}_B^2\rangle-\langle\hat{Q}_B\rangle^2$ can straightforwardly be expressed via the correlation function $C_{mm'} = \langle \hat{\rho}_m \hat{\rho}_{m'}\rangle - \langle \hat{\rho}_m\rangle \langle \hat{\rho}_{m'}\rangle$ by exploiting the exact sum rule $\sum_{m'=1}^{N_s} C_{mm'}=0$. We obtain $\Delta Q_B^2 = -\frac{1}{2} \sum_{m,m'=1}^{N_s} (f_m - f_{m'})^2 C_{mm'}$. Employing that $C_{mm'}$ decays exponentially for $|m-m'|\gg\xi_g$, we find that the fluctuations are finite in the thermodynamic limit and the correlation function $C_{mm'}$ can be replaced by the bulk correlation function $C^{\rm bulk}_{mm'}\equiv a^2 C_{\rm bulk}(x,x')$ as we have $l_p\gg \xi_g$. {\color{black} Expanding $f_m-f_{m'}$ up to first order in $m-m'$ and averaging the correlation function over the unit cell [denoted by $\bar{C}_{\rm bulk}(x,x')=\bar{C}_{\rm bulk}(x-x')$], 
we find in the universal limit $l_p\gg\xi_g\gg a$ the result}
(see the SM \cite{SM} for details)
\begin{align}
   \label{eq:QB_fluc_BBC_main}
   l_p\,\, \Delta Q_B^2 = -\frac{1}{2}\int dx \,x^2 \bar{C}_{\rm bulk}(x) + O(\xi_g^2/l_p)\,,
\end{align}
{\color{black} where we defined $l_p^{-1}=\int dx f'(x)^2$ with $f(ma)\equiv f_m$.} 
Our result requires only the condition that $\bar{C}_{\rm bulk}(x)$ decays exponentially for distances above the scale $\xi_g$. This is expected generically in the insulating regime due to the nearsightedness principle \cite{kohn_prl_96,prodan_kohn_pnas_05}. Here, $\xi_g$ should be considered as an upper limit for the decay length of $\bar{C}_{\rm bulk}(x)$, in multi-channel or interacting models it is generically expected that $\bar{C}_{\rm bulk}(x)$ consists of a linear combination of many exponentially decaying terms with different length scales $\xi_\sigma \lesssim \xi_g$ (see below the discussion of higher-dimensional systems). 

We note that our central result (\ref{eq:QB_fluc_BBC_main}) is independent of the scale $l_p$  (besides the condition $l_p\gg\xi_g$),
offering a high degree of flexibility to measure and calculate the universality of boundary charge fluctuations . E.g., in cold atom systems, one can probe them either directly via the density profile or the correlation function \cite{Bloch_etal_rmp_08}. Alternatively, for {\color{black} the special choice $f_m=1-m/N_s$ (where $l_p=L$), we get $\hat{Q}_B=-\hat{P}/L$ with $\hat{P}=a\sum_{m=1}^{N_s}m(\hat{\rho}_m-N/N_s)$ denoting the bulk polarization operator, the fluctuations of which are also discussed within localization theories \cite{sgiarovello_etal_prb_01}. Our result $l_p\Delta Q_B^2 = \Delta P^2/L$ provides a {\it surface fluctuation theorem} connecting boundary and bulk fluctuations in a universal way. This can be viewed as the fluctuation-based analog of the celebrated surface charge theorem \cite{vanderbilt_kingsmith_prb_93,vanderbilt_book_18,pletyukhov_etal_prbr_20,pletyukhov_etal_prb_20}.} 

\emph{Single-channel case --- }
A theory as general as the one outlined above can be put to the test in a plethora of applications. We start with the most simple single-channel case $N_B=1$ and nearest-neighbor hoppings, where numerically exact results in the clean or disordered case {\color{black}(by diagonalization of the single particle problem)} as well as in the presence of interactions can be obtained with relative ease {\color{black}(in the interacting case by use of density matrix renormalization group approaches)}. We consider the following model 
\begin{align}
    \label{eq:model}
     H= &- \sum_{m=1}^{N_s-1} (t_m + w_m) \big(a_{m+1}^\dagger a_{m} + {\rm h.c.}\big) + \sum_{m=1}^{N_s} v_m \rho_m \notag\\ 
    & + U\sum_{m=1}^{N_s} \big(\rho_m-1/2\big) \big(\rho_{m+1}-1/2\big)\,, 
\end{align}
where $t_m=t_{m+Z}$ and $v_m=v_{m+Z}$ are periodically modulated nearest-neighbor hoppings and on-site potentials, respectively, $w_m$ describes bond disorder drawn from a uniform distribution $w_m\in[-d_m/2,d_m/2)$ with $d_m=d_{m+Z}$, and $U\ge 0$ is a nearest-neighbor repulsive interaction. 

The phase diagram of this model in the SSH limit \cite{su_etal_prl_79,su_etal_prb_80} at half-filling (choosing $Z=2$ and $v_m=0$), and its characterization in terms of the boundary charge fluctuations were already discussed above; see Fig.~\ref{fig:phase_diagram}. Varying the interaction strength $U$ as well as hopping disorder $d=d_1=2d_2$ gap closings indicated by strongly enhanced boundary charge fluctuations are found. In the $(r,d)$-plane at finite disorder and $U=0$ we show that our characterization in terms of the boundary charge fluctuations is perfectly consistent with the number of edge states, thus demonstrating perfect agreement with the theoretical \cite{mondragon-shem_hughes_prl_14} and experimental \cite{meier_etal_science_18} findings.
At finite $U$ the transition to the correlated Mott insulator is more involved and classification schemes are rare. The Mott insulator is characterized by a charge density wave instability due to Umklapp processes \cite{giamarchi_book_03} generating a staggered on-site potential. This potential breaks the chiral symmetry of the SSH model and leads to a non-topological phase. The boundary charge fluctuations provide a valuable tool to find also this transition line; compare Fig.~\ref{fig:phase_diagram}. From exact solutions one point of this transition line into the Mott insulator is known to be at $r=1$ ($t_1=t_2$), $U/t_1=2$, which is in perfect agreement with the boundary charge fluctuation characterization.

\begin{figure}[t]
    \centering
	 \includegraphics[width =\columnwidth]{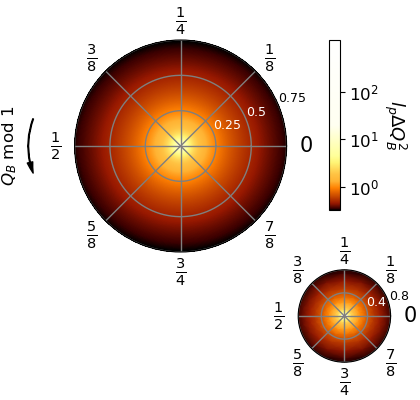}
	  \caption{Polar color plot of $l_p \Delta Q_B^2$ as function of $Q_B$ $\text{mod}(1)$ (polar component) and the gap $E_g$ (radial component) for model (\ref{eq:model}) with $Z=2$, $a=1$, $N/N_s=1/2$, $t=(t_1+t_2)/2=1$, $d=0$, and $U=0$ (upper panel) or $U=0.5$ (lower panel). The data points are obtained by taking $t_{1/2}=t\pm \Delta_0/2 \cos(\phi)$, $v_{1/2}=\pm \Delta_0 \sin(\phi)$, and varying the parameters $\Delta_0$ and $\phi$ in the intervals $\Delta_0 \in [0, 0.375]$ and $\phi\in[0,2\pi]$. For $U=0$ the analytic result ($N_s,l_p\to\infty$) is used, while for $U=0.5$  we set $N_s=1000$, and $l_p=400$.   
	  }
    \label{fig:polar_plot}
\end{figure}

Whereas the boundary charge fluctuations depend strongly on the gap size, the boundary charge $Q_B$ itself is sensitive to the phase of the gap parameter in one-dimensional, single-channel models \cite{pletyukhov_etal_prr_20}. This suggests the polar plot of Fig.~\ref{fig:polar_plot}, where we show the fluctuations $l_p \Delta Q_B^2$ in dependence of the gap $E_g$ (radial component) and the boundary charge $Q_B$ $\text{mod}(1)$ (polar component) for the model of Eq.~(\ref{eq:model}) with $Z=2$ and $d=0$, both for $U=0$ and finite $U$, choosing a variety of parameters to define the staggered on-site potentials $v_1=-v_2$ and hoppings $t_{1/2}$. This corresponds to the noninteracting and interacting Rice-Mele model \cite{rice_mele_prl_82,lin_etal_prb_20}. The radially symmetric value of $l_p \Delta Q_B^2$ indicates that the fluctuations depend only on the gap's absolute value but not on $Q_B$ and that they strongly enhance at the center $E_g\to0$. We expect this feature to be  generic for one-dimensional, single-channel models in the low-energy regime.

\begin{figure}[t]
    \centering
	 \includegraphics[width =\columnwidth]{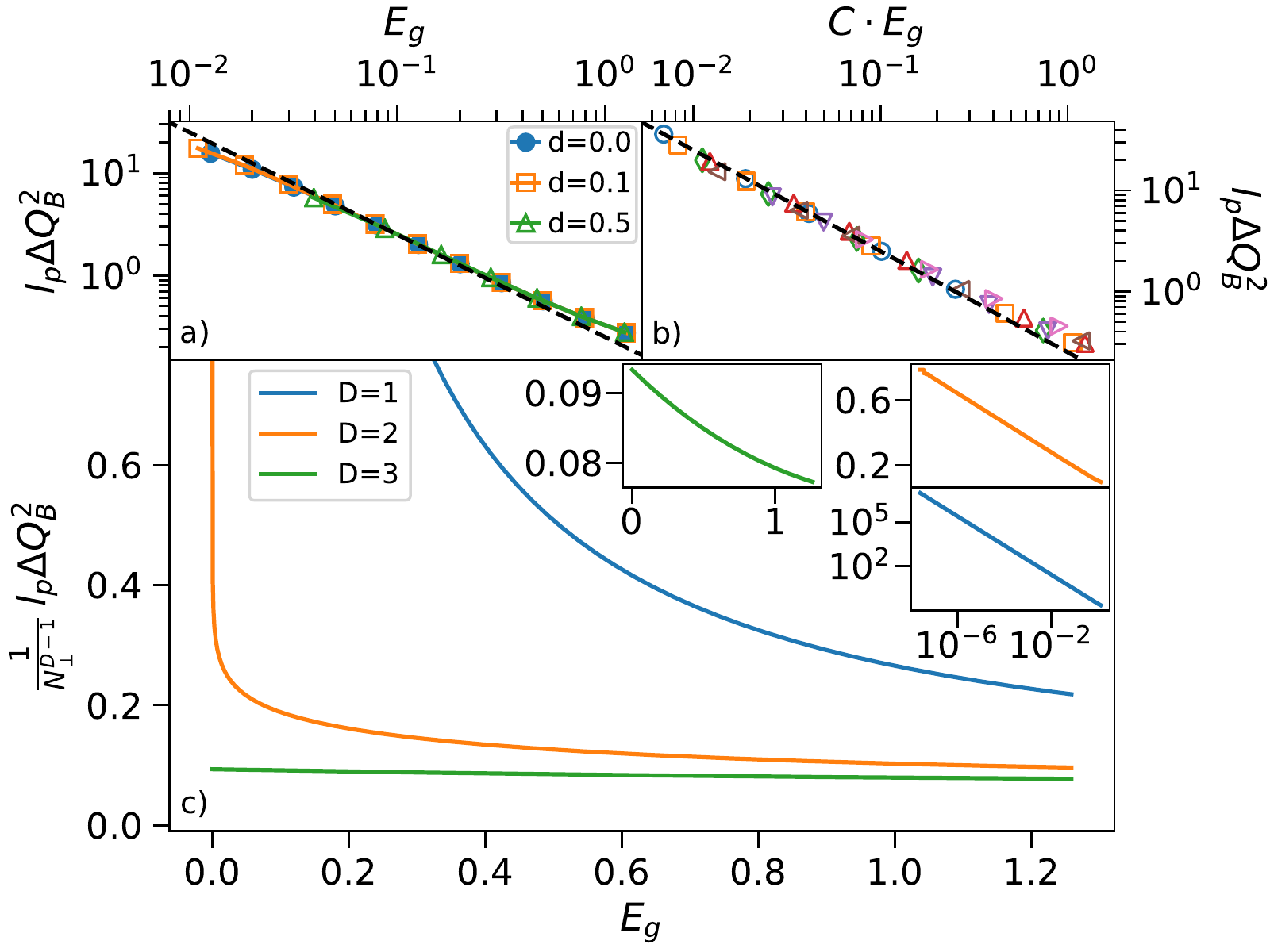}
	  \caption{(a-b) Scaling of $l_p \Delta Q_B^2$ with the gap $E_g$ for model (\ref{eq:model}) on double-logarithmic scale for (a) $U=0$ and varying disorder strength $d$ (averaged over 20 disorder configurations) and (b) $d=0$ and varying Coulomb interaction strength $U$ (other parameters as in Fig.~\ref{fig:phase_diagram}). The dashed line indicates $v_{\rm F}/(8 E_g)=1/(4 E_g)$. In (b) $U=(0.0, 0.4, 0.8, 1.2, 1.6, 2.0, 2.4)$ are given by (blue, orange, green, red, purple, brown, pink) symbols. The non-universal perfactors $C(U)$ for the collapse are given in the SM \cite{SM}. In panels (a) and (b) we choose $N_s=1000$ and $l_p=400$. (c) Scaling of $l_p \Delta Q_B^2/N_\perp^{D-1}$ from the analytic result ($N_s,l_p\to\infty$) with the gap $E_g=2|t_1-t_2|$ for various dimensions $D$ for the SSH model ($D=1$) and for the higher-dimensional models discussed in the main text ($D=2,3$) with $t=(t_1+t_2)/2=t_y=t_z=1$. The insets show the same results on different scales.}
    \label{fig:scaling}
\end{figure}

For the noninteracting and clean Rice-Mele model $Z=2$ we find analytically the exact result (see the SM for details \cite{SM})
$l_p \, \Delta Q_B^2 = a(t_1^2 + t_2^2)/\big(4 E_g\sqrt{E_g^2/4 + 4 t_1 t_2}\big)$, with the gap $E_g=2\sqrt{v^2 + (t_1-t_2)^2}$ and $v=v_1=-v_2$. In the vicinity of phase transitions $E_g\ll t=(t_1+t_2)/2$, we obtain the universal scaling
\begin{align}
    \label{eq:universal_scaling}
    l_p \, \Delta Q_B^2(E_g) \xrightarrow{E_g\ll t} \frac{v_F}{8 E_g} = \frac{\xi_g}{8} \,,
\end{align}
where $v_F=2ta$ denotes the Fermi velocity. For arbitrary value of $Z$ and generic modulations of the nearest-neighbor hoppings and the on-site potentials we confirm this universal scaling for the chemical potential located in any gap. We obtain this result by using the exact eigenstates of a low-energy massive Dirac model in $1+1$ dimensions, as proposed in Ref.~\cite{pletyukhov_etal_prr_20} (see the SM \cite{SM} for details). Furthermore, we find that $\xi_g=v_F/E_g$ is the exponential decay length of the correlation function $\bar{C}_{\rm bulk}(x)$. For more exotic models which can not be described by a Dirac model in the low-energy regime, we show in the SM \cite{SM} that also other scalings are in principle possible. For the noninteracting and clean SSH model, we numerically confirm the scaling $l_p \Delta Q_B^2=v_F/(8 E_g)$ in Fig.~\ref{fig:scaling} a) and find that it holds up to surprisingly large gaps even beyond the applicability range of the low-energy theory. This relation holds also in the presence of disorder, at least for not too strong disorder $d\lesssim 2-3$, where a Born approximation \cite{groth_etal_prl_09} can be used to define a renormalized gap $E_g=2|\bar{t}_1-\bar{t}_2|$, with $\bar{t}_1=t_1-\theta(t_1-t_2)d_1^2/(12t_1)$ and $\bar{t}_2=t_2-\theta(t_2-t_1)d_2^2/(12 t_2)$ denoting renormalized hoppings (see the SM for details \cite{SM}). In Fig.~\ref{fig:scaling} b) we show the scaling for different $U$ and $d=0$ and find that they collapse to the universal $\sim 1/E_g$ if one allows for a $U$ dependent non-universal prefactor \cite{SM}. In this case the gap is significantly increased by interactions \cite{kivelson_etal_prb_85,Horovitz_Solyom_prb_85,gangadharaiah_etal_prl_12,pletyukhov_etal_prr_20,lin_etal_prb_20}. 

\emph{Two- and Three-Dimensional Systems --- }
To analyze the scaling in two and three dimensions $D=2,3$ for noninteracting and clean systems we use translational invariance in the transverse direction and consider $N_B=N_\perp^D$ transverse quasimomenta $\underline{k}_\perp$ as a channel index. The fluctuations of $Q_B$ can then be calculated as an independent sum $\Delta Q_B^2 = \sum_{\underline{k}_\perp}\Delta Q_B^2(\underline{k}_\perp)$, with $k_\perp=k_y$ for $D=2$ and $\underline{k}_\perp=(k_y,k_z)$ for $D=3$. For each fixed $\underline{k}_\perp$, we consider an effective one-dimensional, single-channel system and get from Eq.~(\ref{eq:universal_scaling}) in the low-energy regime $l_p \Delta Q_B^2(\underline{k}_\perp)=v_F(\underline{k}_\perp)/(8E_g(\underline{k}_\perp))$, corresponding via Eq.~(\ref{eq:QB_fluc_BBC_main}) to an independent term of $\bar{C}_{\rm bulk}(x)$ decaying on length scale $\xi(\underline{k}_\perp)=v_F(\underline{k}_\perp)/E_g(\underline{k}_\perp) \le \xi_g$. The momentum dependence of the effective gap $E_g(\underline{k}_\perp)$ can be estimated for a typical massive Dirac-like spectrum in $D+1$ dimensions: $E_g(\underline{k}_\perp)\approx 2\sqrt{\bar{v}_F^2 k_\perp^2 + E_g^2/4}$, where $E_g$ denotes the spectral gap, and we have neglected the weak momentum dependence of $v_F(\underline{k}_\perp)\approx \bar{v}_F$. In the thermodynamic limit $\sum_{\underline{k}_\perp}\rightarrow \big(N_\perp a/(2\pi)\big)^{D-1}\int_{-\pi/a}^{\pi/a} d^{D-1}k_\perp$, we can estimate the scaling of the fluctuations. In two dimensions, we obtain a logarithmic scaling $N_\perp^{-1}l_p \Delta Q_B^2\sim(\bar{v}_F/W)\ln(W/E_g)\sim a\ln(\xi_g/a)$, where $W$ defines a high-energy cutoff scale for $|\bar{v}_F k_y|$. In contrast, for three dimensions we obtain a monotonic increase for the fluctuations with decreasing gap but a finite value in the zero-gap limit. 

Systems illustrating this generic behavior can be realized, e.g., in cold atom systems \cite{cooper_etal_rmp_19}. As an example we consider an SSH model in $x$-direction (with alternating hoppings $t_{1,2}$), constant nearest-neighbor hoppings in transverse direction (denoted by $t_{y,z}$), and a homogeneous magnetic field of size $B$ in $z$-direction (for $D=2$) or in the $y$- and $z$-direction (for $D=3$). For the simplest case that the magnetic length is given by $\lambda_B=2a$, we obtain in the Landau gauge an effective one-dimensional Rice-Mele model with $E_g(\underline{k}_\perp)=2\sqrt{v(\underline{k}_\perp)^2+(t_1-t_2)^2}$, where $v(\underline{k}_\perp)=2t_y \cos(k_y a)$ in $D=2$ or $v(\underline{k}_\perp)=2t_y \cos(k_y a)+2t_z \cos(k_z a)$ in $D=3$ (see the SM \cite{SM} for details). Using the exact result for the Rice-Mele model to calculate $l_p\Delta Q_B^2(\underline{k}_\perp)$, one can perform the integral over $\underline{k}_\perp$, and finds for the fluctuations as function of the gap $E_g=2|t_1-t_2|$ the result shown in Fig.~\ref{fig:scaling}~c). The logarithmic scaling in $D=2$ is perfectly preserved even for large gaps, suggesting the boundary charge fluctuations to be also useful as an indicator for phase transitions in two dimensions. {\color{black} We emphasize that for $D=3$, we observe only a weak increase of the fluctuations with a finite value at zero gap. Therefore, for three dimensions, the fluctuations are only a weak indicator for the transition. Although this limits the universally diverging behavior to one- and two-dimensional systems, future studies should address whether also the non-divergent enhancement caries valuable information of the type of transition passed. In addition, it will of interest to study how generic the proposed decoupling in transverse modes will persist in higher-dimensional interacting and disordered systems.}

\emph{Conclusion --- }
We have  established the boundary charge and its fluctuations as a measurable tool to probe topological properties of insulators. {\color{black} Whereas the boundary charge takes the role of a phase and jumps by $e/2$ at a topological phase transition \cite{pletyukhov_etal_prr_20}, the complementary fluctuations are strongly enhanced in one and two dimensions and reveal a universal scaling as function of the gap size. In contrast to the number of topological edge states which is controlled by non-fluctuating topological indices, we found that the universal scaling properties of the fluctuations do not depend on whether a topological index changes at the transition but rely exclusively on the characteristic band structure of insulators with topological properties.} Importantly, this characterization scheme can be applied to band, Anderson, and Mott insulators alike. {\color{black} An intriguing avenue of future research concerns the question whether the characterization proposed here is also useful in the context of topological superconductors, for which simple models do not fulfill charge conservation. }

\emph{Acknowledgments --- }
We thank V. Meden and D. Schuricht for fruitful discussions. This work was supported by the Deutsche Forschungsgemeinschaft via RTG 1995, the Swiss National Science Foundation (SNSF) and NCCR QSIT and by the Deutsche Forschungsgemeinschaft (DFG, German Research Foundation) under Germany's Excellence Strategy - Cluster of Excellence Matter and Light for Quantum Computing (ML4Q) EXC 2004/1 - 390534769. We acknowledge support from the Max Planck-New York City Center for Non-Equilibrium Quantum Phenomena. Simulations were performed with computing resources granted by RWTH Aachen University under project thes0753. Funding was received from the European Union's Horizon 2020 research and innovation program (ERC Starting Grant, grant agreement No 757725).

C.S.W., K.P. and M.P. contributed equally to this work.

\bibliography{citations_3}

\begin{thebibliography}{53}
\bibitem [{Fig()}]{Fig1_para}%
  \BibitemOpen
  \href@noop {} {\ }\bibinfo {note} {\hspace{-0.125cm}The other parameters are
  $N_s=400$, $l_p=200$ for $d\neq 0$ and $N_s,l_p\to\infty$ for $d=0$ (see SM
  \cite{SM} for details how the infinite system size limit can be taken
  numerically), $t_2=1$, $t_1=r$, $a=1$ and we work at half-filling. The
  results are averaged over 20 disorder configurations. For the details of the
  model see Eq. \eqref{eq:model}}\BibitemShut {NoStop}%
\end{thebibliography}

\end{document}


\title{Universality of Boundary Charge Fluctuations: Supplemental Material}

\author{Clara S. Weber}
\affiliation{Institut f\"ur Theorie der Statistischen Physik, RWTH Aachen, 
52056 Aachen, Germany and JARA - Fundamentals of Future Information Technology}

\author{Kiryl Piasotski}
\affiliation{Institut f\"ur Theorie der Statistischen Physik, RWTH Aachen, 
52056 Aachen, Germany and JARA - Fundamentals of Future Information Technology}

\author{Mikhail Pletyukhov}
\affiliation{Institut f\"ur Theorie der Statistischen Physik, RWTH Aachen, 
52056 Aachen, Germany and JARA - Fundamentals of Future Information Technology}

\author{Jelena Klinovaja}
\affiliation{Department of Physics, University of Basel, Klingelbergstrasse 82, 
CH-4056 Basel, Switzerland}
\author{Daniel Loss}
\affiliation{Department of Physics, University of Basel, Klingelbergstrasse 82, 
CH-4056 Basel, Switzerland}
\author{Herbert Schoeller}
\affiliation{Institut f\"ur Theorie der Statistischen Physik, RWTH Aachen, 
52056 Aachen, Germany and JARA - Fundamentals of Future Information Technology}
\author{Dante M. Kennes}
\email[Email: ]{Dante.Kennes@rwth-aachen.de}
\affiliation{Institut f\"ur Theorie der Statistischen Physik, RWTH Aachen, 
52056 Aachen, Germany and JARA - Fundamentals of Future Information Technology}
\affiliation{Max Planck Institute for the Structure and Dynamics of Matter, Center for Free Electron Laser Science, 22761 Hamburg, Germany}


\maketitle

\section{Numerical results of Fig.4b}
\label{sec:num_result}

\begin{table}[h]
\begin{tabular}{p{0.5cm}
>{\raggedleft\arraybackslash}p{0.8cm}
>{\raggedleft\arraybackslash}p{0.8cm}
>{\raggedleft\arraybackslash}p{0.8cm}
>{\raggedleft\arraybackslash}p{0.8cm}
>{\raggedleft\arraybackslash}p{0.8cm}
>{\raggedleft\arraybackslash}p{0.8cm}
>{\raggedleft\arraybackslash}p{0.8cm}}
\hline
$U$ &  $0.0$ & $0.4$   & $0.8$   & $1.2$ & $1.6$ & $2.0$  & $2.4$   \\ \hline
$C$  & $1.0$ & $0.92$ & $0.88$ & $0.96$ & $0.89$ &$0.96$ & $1.18$ \\ \hline
\end{tabular}
\caption{Constants $C$ used in the collapse of Fig.~4 (b) of the main text. }
\label{tab:const}
\end{table}

To achieve the universal collapse shown in Fig.~4 (b) of the main text we have numerically determined the constants $C$ listed in Table \ref{tab:const}.

\section{Boundary charge fluctuations}
\label{sec:fluctuations}

Here we present a general analysis of the boundary charge fluctuations without assuming that the gap is small compared to the band width. This means that the length scale $\xi_g=v_F/E_g$ is not assumed to be much larger than the lattice spacing $a$. As in the main part of the letter we assume fixed particle number $N$ (i.e., a canonical ensemble) and zero temperature here (other cases are discussed in Section~\ref{sec:finite_T}). We take an effectively one-dimensional system with $N_s$ lattice sites and $N_B$ channels per site, labeled by $m=1,\dots,N_s$ and $\sigma=1,\dots,N_B$, respectively. Besides different spins, orbitals and other flavors, the channel indizes include also the degrees of freedom in transverse direction, i.e., parallel to the surface. The size of the one-dimensional unit cell is denoted by $Za$. For the real space position of a lattice site we write $x=ma$, and $L=N_s a$ defines the system size perpendicular to the boundary. Finally, we take the thermodynamic limit $N_s,N\rightarrow\infty$ such that the average charge per site $\bar{\rho}=N/N_s$ is kept constant. We use units $\hbar=e=1$.

In Section~\ref{sec:correlation} we analyse the relation of the boundary charge fluctuations to the second moment of the longitudinal density-density correlation function. In Section~\ref{sec:finite_T} we discuss the case of a grandcanonical ensemble and finite temperature.

\subsection{Relation to the density-density correlation function}
\label{sec:correlation}

Our aim is to calculate the fluctuations of the boundary charge operator at one end of the system defined by
\begin{align}
    \label{eq:QB_operator}
    \hat{Q}_B = \sum_{m=1}^{N_s} f_m (\hat{\rho}_m - \bar{\rho})\,,
\end{align}
where $\hat{\rho}_m=\sum_\sigma a^\dagger_{m\sigma}a_{m\sigma}$ is the charge operator at site $m$, and $a_{m\sigma}^\dagger$ creates a fermion on site $m$ in channel $\sigma$. The macroscopic average is described by an envelope function $f_m\equiv f(x)=1-\theta_{l_p}(x-L_p)$, with $L\gtrsim L_p + l_p/2 \gtrsim l_p\gg \xi_g$, where $\theta_{\delta x}(x)$ is some representation of the $\theta$-function broadened with $\delta x$. The scale $L_p$ describes the length of a charge measurement probe and $l_p$ is the scale on which the probe smoothly looses the contact to the sample. By convention, we define the scale $l_p$ by the integral
\begin{align}
    \label{eq:delta_l_p}
    l_p^{-1} = \int dx [f'(x)]^2 \,.
\end{align}
The envelope function is assumed to be smooth on the microscopic length scales $\xi_g$ and $a$, i.e., $l_p\gg \xi_g,a$. As shown below the length scales $L$, $L_p$, and $l_p$ can even be of the same order of magnitude, provided that $|L-L_p-l_p/2|\gtrsim O(\xi_g)$ and $|L_p-l_p/2|\gtrsim O(\xi_g)$. This condition means that the fall-off of the envelope function fits into the system size (up to $O(\xi_g)$). Otherwise, as we will see below, the length scale $L_p$ does not enter the final solution for the boundary charge fluctuations. 

Defining the correlation function  
\begin{align}
    \label{eq:C}
    C_{mm'} = \langle \hat{\rho}_m \hat{\rho}_{m'}\rangle - \langle \hat{\rho}_m\rangle \langle \hat{\rho}_{m'}\rangle\,, 
\end{align}
we can express the fluctuations $\Delta Q_B^2=\langle \hat{Q}_B^2\rangle - {\langle \hat{Q}_B\rangle}^2$ as 
\begin{align}
    \label{eq:QB_fluc_C_zw}
    \Delta Q_B^2 = \sum_{m,m'=1}^{N_s} f_m f_{m'} C_{mm'}\,.
\end{align}
In the following we call $C_{mm'}$ the density-density correlation function of the effectively one-dimensional system but it should be kept in mind that it is the correlation between the charges $\rho_m$ and $\rho_{m'}$ including the sum over {\it all} channel indizes. In particular, this includes also the sum over all transverse quasimomenta. Therefore, for higher-dimensional systems, it describes rather the correlation between the total charge of two stripes and not the correlation between the densities at two points in real space.

Using the sum rule (which is exact at fixed particle number) 
\begin{align}
    \label{eq:sum_rule}
    \sum_{m'=1}^{N_s} C_{mm'}=0\,,
\end{align} 
together with $C_{mm'}=C_{m'm}$, we can replace $f_m f_{m'}\rightarrow -(1/2)(f_m - f_{m'})^2$ in (\ref{eq:QB_fluc_C_zw}) and obtain the very useful form
\begin{align}
    \label{eq:QB_fluc_K}
    \Delta Q_B^2 = -\frac{1}{2} \sum_{m,m'=1}^{N_s} (f_m - f_{m'})^2 C_{mm'}\,.
\end{align}
This formula is very helpful since the correlation function $C_{mm'}$ is exponentially small for an insulator for $|x-x'|\gg\xi_g$, with $x=ma$ and $x'=m'a$. Therefore, only the part $|x-x'|\lesssim\xi_g$ is relevant. Since $f_m\equiv f(x)$ varies slowly on the scale $\xi_g$ due to $\xi_g\ll l_p$, we can expand $f_m-f_{m'}\approx f'(\frac{x+x'}{2})(x-x')$ in (\ref{eq:QB_fluc_K}). Using in addition that the derivative $f'(x)\sim 1/{l_p}$ is only non-zero for $x=L_p + O(l_p/2)$, we find that both $x,x'\sim L_p + O(l_p/2)\gg\xi_g$ are located away from the boundary far beyond the length scale $\xi_g$. This holds even in the case when $L_p$ and $l_p/2$ have the same order of magnitude (up to $O(\xi_g)$) since the contribution from all $x,x'\sim\xi_g$ gives a contribution $\lesssim\big(\frac{\xi_g}{l_p}\big)^2$ to the fluctuations. Therefore, we can replace the correlation function $C_{mm'}$ by its bulk value
\begin{align}
    \label{eq:C_bulk_def}
    C^{\rm bulk}_{mm'} = a^2 C_{\rm bulk}(x,x')\,, 
\end{align}
and obtain for $l_p\gg \xi_g$
\begin{align}
    \nonumber
    \Delta Q_B^2 &= -\frac{1}{2}a^2\sum_{m,m'=1}^{N_s} \left[f'(\frac{x+x'}{2})\right]^2\\
    \label{eq:QB_fluc_bulk}
    &\hspace{1cm} \times (x-x')^2 C_{\rm bulk}(x,x') + O(\frac{\xi_g}{l_p})^2\,.
\end{align}
As a result, the double sum scales as $a^2\sum_{m,m'}\lesssim \xi_g l_p$ and one can see that the fluctuations are finite in the thermodynamic limit. In this formula the bulk correlation function can be calculated in the thermodynamic limit for any ensemble and for any boundary condition. Note that this is not possible in the form (\ref{eq:QB_fluc_C_zw}) since the sum rule (\ref{eq:sum_rule}) does no longer hold exactly in a grandcanonical ensemble at finite temperature (see the discussion in Section~\ref{sec:finite_T}).

Using $m=Z(n-1)+j$, where $j=1,\dots,Z$ denotes the site index within a unit cell and $n=1,2,\dots$ labels the unit cells, we can write $C_{\rm bulk}(x,x')=C^{\rm bulk}_{jj'}(Zna,Zn'a)$. Due to translational invariance on the size $Za$ of a unit cell (which holds on average also in the presence of random disorder) the bulk correlation function depends only on the difference $n-n'$ and we can write
\begin{align}
    \label{eq:C_bulk_relative}
    C_{\rm bulk}(x,x') = C^{\rm bulk}_{jj'}\big(Z(n-n')a\big)\,. 
\end{align}
Using $f'(\frac{x+x'}{2})=f'(Z n_s a)(1+O(a/l_p)$, with $n_s=\frac{n+n'}{2}$, we find that the sum over $n_s$ in (\ref{eq:QB_fluc_bulk}) gives $\sum_{n_s} f'(Z n_s a)^2=\frac{1}{Za l_p}\big(1+O(Za/l_p)\big)$ according to (\ref{eq:delta_l_p}). Again neglecting boundary effects $\sim O(\frac{\xi_g}{l_p})^2$, we obtain 
\begin{align}
    \nonumber 
    l_p \, \Delta Q_B^2 &= -\frac{a}{2Z}\sum_n \sum_{jj'}\\
    \label{eq:QB_fluc_BBC}
    &\hspace{-0.5cm}
    \times\big(Zna + (j-j')a\big)^2 C^{\rm bulk}_{jj'}(Zna) \,.
\end{align} 
We note that this result is exact when performing the limits $L,L_p,l_p\rightarrow\infty$ with $L\gtrsim L_p + l_p/2 \gtrsim l_p$. Eq.~(\ref{eq:QB_fluc_BBC}) establishes  a universal relation of the boundary charge fluctuations to the density-density correlation function of the bulk, including all microscopic details of the unit cell. As one can see only the product $l_p \, \Delta Q_B^2$ is related in a universal way to the bulk correlation function $C_{jj'}^{\rm bulk}(Zna)$, where $l_p$ is defined by Eq.~(\ref{eq:delta_l_p}) in terms of the envelope function. Since the bulk correlation function can be calculated in any ensemble and with any boundary condition this equation is the most convenient starting point to calculate the boundary charge fluctuations very efficiently from bulk quantities without any need to determine the complicated eigenfunctions of finite or half-infinite systems.  

As one can see from the proof only the condition $l_p\gg \xi_g$ enters together with the property that the support of the function $f'(x)$ fits into  the system size such that 
\begin{align}
    \label{eq:f_der_integral}
-\int_0^L dx f'(x) &= 1+O(\xi_g/l_p) \,,\\
\label{eq:f_der_squared_integral}
\int_0^L dx \Big[f'(x)\Big]^2 &= l_p^{-1} \Big(1+O(\xi_g/l_p)\Big)\,.
\end{align}
This is the reason why the length scales $L$, $L_p$, $l_p$ can all be of the same order of magnitude (in the sense defined above) without changing the leading order contribution of the fluctuations. This is very helpful for numerical calculations and the experimental observability since only the condition $l_p\gg\xi_g\gg a$ has to be fulfilled to find universal properties of the boundary charge fluctuations. 

To proceed we note at this point that all universal properties of $\Delta Q_B^2$ derived in this section follow from two fundamental properties of $C^{\rm bulk}_{jj'}(x)$ which have to be checked for the concrete model under consideration
\begin{align}
    \label{eq:C_exp}
    C^{\rm bulk}_{jj'}(x) &= \sum_\sigma \frac{1}{\xi_\sigma^2} \,g^{(\sigma)}_{jj'}(x/\xi_\sigma) e^{-x/\xi_\sigma} \,,\\
    \label{eq:C_scale}
    s^2 g^{(\sigma)}_{jj'}(s) &\sim O(1) \quad {\rm for} \quad |s|<1\,.
\end{align}
The first condition (\ref{eq:C_exp}) states exponential decay and indicates that the correlation function is in general a linear combination of many terms, each with its own decay length $\xi_\sigma < \xi_g$. This occurs generically in the presence of channel indizes describing flavor degrees of freedom (spin, orbital, etc.) or the transverse quasimomentum (see Section~\ref{sec:quantum_hall_insulator}). Due to the second condition (\ref{eq:C_scale}) the pre-exponential function should scale as $g^{(\sigma)}_{jj'}(s)\sim 1/s^2$ for $|s|<1$. We find this scaling independent of the microscopic details of the model. For $\xi_g\sim a$ this property is obvious since there is only a single length scale. In the low-energy regime $\xi_g\gg a$, we show in Section~\ref{sec:C_dirac_noninteracting} explicitly that the two properties are fulfilled for single-channel and noninteracting models. The physical reason for the general case is obvious. For $|x|\sim a\ll \xi_\sigma < \xi_g$ one probes high-energy scales where the gap is unimportant. Therefore, the correlation function will scale $\sim 1/a^2$. For $|x|\sim \xi_\sigma \gg a$, the lattice spacing does not play any role and a low-energy continuum theory is possible to describe the corresponding term of the correlation function. This theory is expected to depend mainly on a single length scale $\xi_\sigma$, such that the corresponding term of the correlation function will scale $\sim 1/\xi_\sigma^2$. In contrast, for $|x|\gg \xi_\sigma$ or $|s|\gg 1$, the scaling depends crucially on the low-energy properties of the model and a pre-exponential power-law with an interaction dependent exponent is expected. The latter is difficult to determine for interacting systems. For clean single-channel systems one obtains $1/|s|$ in the noninteracting case, see Section~\ref{sec:noninteracting}. However, this regime is of no relevance for the fluctuations since the corresponding term of the correlation function is exponentially small for $|x|\gg\xi_\sigma$. We note that the exponential decay property (\ref{eq:C_exp}) is also valid for the correlation function $C_{mm'}$ with a boundary but the scaling of the pre-exponential function for $n,n'$ close to the boundary might be more subtle.

Using (\ref{eq:C_exp}) and (\ref{eq:C_scale}) one can estimate the order of magnitude of the fluctuations (\ref{eq:QB_fluc_BBC}) as
\begin{align}
    \label{eq:QB_fluc_order}
    l_p \, \Delta Q_B^2 = \sum_\sigma c_\sigma\xi_\sigma \equiv N_B\bar{\xi} \lesssim N_B \xi_g \,,
\end{align}
with $c_\sigma\sim O(1)$, and $\bar{\xi}=\frac{1}{N_B}\sum_\sigma c_\sigma\xi_\sigma\le\xi_g$ defining some average exponential decay length. This result shows that the boundary charge fluctuations $\Delta Q_B\lesssim \sqrt{N_B \xi_g/l_p}\ll \sqrt{N_B}$ are always much smaller than the boundary charge $Q_B\sim N_B$, even for $N_B\sim O(1)$, showing that the boundary charge is a well-defined observable for $l_p\gg\xi_g$ \cite{park_etal_prb_16}, in analogy to interface charges studied in Refs.~[\onlinecite{kivelson_schrieffer_prb_82,bell_rajaraman_plb_82,kivelson_prb_82,bell_rajaraman_npb_83,frishman_horovitz_prb_83,jackiw_etal_npb_83}].

In the low-energy regime when $\xi_g\gg Za$, one can neglect all terms in (\ref{eq:QB_fluc_order}) with $\xi_\sigma\sim Za\ll \xi_g$. For the terms with $\xi_\sigma\gg Za$, we can neglect the part $(j-j')a\sim Za\ll \xi_\sigma \sim Zna$ in (\ref{eq:QB_fluc_bulk}) and the sum can be replaced by an integral. This leads to the compact formula
\begin{align}
    \label{eq:QB_fluc_BBC_low_energy}
    l_p \, \Delta Q_B^2 = -\frac{1}{2}\int dx \,x^2 \bar{C}_{\rm bulk}(x)\,,
\end{align} 
where $\bar{C}_{\rm bulk}(x)$ is the correlation function averaged over $j$ and $j'$ 
\begin{align}
    \label{eq:C_n}
    \bar{C}_{\rm bulk}(x)=\frac{1}{Z^2}\sum_{jj'}C^{\rm bulk}_{jj'}(x)\,.
\end{align}

\subsection{Numerics in the infinite system size limit $N_s\to\infty$ }
One of the many important implications of the previous subsection is that the boundary charge fluctuations $l_p \, \Delta Q_B^2 $ can be extracted either directly or equivalently (in the $L_p,l_p\to \infty$ limit) from the right hand side of Eq.~\eqref{eq:QB_fluc_BBC} via the bulk correlation functions. In the translationally invariant case this can be utilized to significantly speed up the numerical determination of  $l_p \, \Delta Q_B^2 $ as one can work directly in the desired limit of $N_s\to\infty$, which is often more convenient. We use this in Fig.~1 of the main text to determine the non-zero $U$, but $d=0$ data from a highly efficient infinite system size density matrix renormalization group approach. With this the problem is directly phrased in the correct limit of $N_s,L_p,l_p\to\infty$, with $N_s a\gtrsim L_p + l_p/2 \gtrsim l_p \gg \xi_g$, but the evaluation of  Eq.~\eqref{eq:QB_fluc_BBC} requires the evaluation of an infinite sum $\sum_{n}$. As usual in numerical approaches we truncate this infinite sum at finite, but large index by $\sum_{n=-\infty}^{\infty}\to \sum_{n=-n_c}^{n_c}$, where $n_c$ needs to be converged to $n_c a\gg \xi_g$ (akin, but not exactly equal to keeping a finite $l_p$) . We choose $n_c=499$ for the data shown in Fig.~1 of the main text.

\subsection{Finite temperature and grandcanonical ensemble}
\label{sec:finite_T}

For a canonical ensemble at fixed particle number, all our results are also valid at finite temperature $T$, provided that $T\ll\Delta$ is much smaller than the gap $E_g = 2\Delta$. In this case the finite temperature corrections to $\Delta Q_B^2$ can be shown to be exponentially small of relative order $\sim \sqrt{T/\Delta}\, e^{-\Delta/T}$, see Section~\ref{sec:rice_mele_finite_T}. This changes for a grandcanonical ensemble where the sum rule (\ref{eq:sum_rule}) is no longer fulfilled exactly. For an estimation we consider a non-interacting system and find after a straightforward calculation using Wick's theorem
\begin{align}
    \nonumber 
    \sum_{m'=1}^{N_s} C_{mm'} =& \\
    \nonumber &
    \hspace{-1cm}
    =\sum_s n_F(\epsilon_s) [1-n_F(\epsilon_s)] \sum_\sigma |\psi_s(m\sigma)|^2 \\
    \label{eq:sum_rule_grand}
    &\hspace{-1cm}
    \sim N_B\frac{T}{W}\,e^{-\Delta/T}\,,
\end{align}
where $\psi_s(m\sigma)$ are the single-particle eigenstates with energy $\epsilon_s$, $n_F(\epsilon_s)$ denotes the Fermi function, and $W$ is the band width. As a result we find that in the steps to get Eq.~(\ref{eq:QB_fluc_K}) from (\ref{eq:QB_fluc_C_zw}), the contributions from $f_m^2$ and $f_{m'}^2$ lead to corrections of order 
\begin{align}
    \label{eq:QB_fluc_grand_cor}
    \Delta Q_B^2(T) - \Delta Q_B^2(T=0) \sim N_B\frac{L_p}{a}\frac{T}{W}e^{-\Delta/T}\,. 
\end{align}
With $\Delta Q_B^2(T=0)\sim N_B\frac{\bar{\xi}}{l_p}$, and the thermal length $L_T\sim v_F/T\sim aW/T$ we conclude that the temperature dependent correction is of relative order 
\begin{align}
    \label{eq:QB_fluc_relative_T}
    \frac{\Delta Q_B^2(T) - \Delta Q_B^2(T=0)}{\Delta Q_B^2(T=0)} \sim \frac{L_p l_p}{\bar{\xi}^2}\frac{\bar{\xi}}{L_T} e^{-\Delta/T}\,. 
\end{align}
However, even though $L_p l_p \gg \xi_g^2 > \bar{\xi}^2$, we expect these corrections to be very small at low temperatures $\bar{\xi}<\xi_g\ll L_T$ due to the exponentially small factor $e^{-\Delta/T}$. As a consequence we conclude that all our central results remain valid for a grandcanonical ensemble as well.

\section{Noninteracting and clean models}
\label{sec:noninteracting}

In this section we analyse the special case of noninteracting and clean systems. In Section~\ref{sec:C_properties} we provide general reasons for the properties (\ref{eq:C_exp}) and (\ref{eq:C_scale}) by expressing the correlation function via the propagator. For the special case of the Rice-Mele model we present the exact solution for the fluctuations of the boundary charge in Section~\ref{sec:rice_mele}. Finally, in Section~\ref{sec:C_dirac_noninteracting} we present the generic low-energy theory for all single channel models in the low-energy regime in terms of a noninteracting Dirac model following the ideas of Ref.~\onlinecite{pletyukhov_etal_prr_20}.

\subsection{Properties of the density-density correlation function and the propagator}
\label{sec:C_properties}

Here we provide qualitative reasons why the properties (\ref{eq:C_exp}) and (\ref{eq:C_scale}) are fulfilled. Due to translational invariance perpendicular to the effectively one-dimensional system (i.e., parallel to the boundary), we consider in the following a fixed value for the transverse quasimomentum, restricting the sum over the channels only to a finite and small number $N_c$ of other flavor degree of freedom. 

For any noninteracting and clean lattice model in a grandcanonical ensemble one can use Wick's theorem and obtains for the density-density correlation function
\begin{align}
    \label{eq:C_U=0_clean}
    C^{\rm bulk}_{mm'} = \delta_{mm'}\rho_m - \sum_{\sigma,\sigma'=1}^{N_c} |\langle a^\dagger_{m\sigma}a_{m'\sigma'}\rangle|^2\,.
\end{align}
The propagator $\langle a^\dagger_{m\sigma}a_{m'\sigma'}\rangle$ can be expressed via the single-particle Bloch eigenfunctions 
\begin{align}
    \label{eq:Bloch_eigenstates}
    \psi_k^{(\alpha)}(m,\sigma) = \frac{1}{\sqrt{2\pi}} \chi_k^{(\alpha)}(j,\sigma) e^{ikn}\,,
\end{align}
with energy $\epsilon_k^{(\alpha)}$, where $\alpha$ is the band index and $-\pi < k < \pi$ denotes the quasimomentum in units of the inverse lattice spacing. With $a^\dagger_{m\sigma}=\sum_\alpha\int_{-\pi}^\pi dk \,c^\dagger_{k\alpha}\psi_k^{(\alpha)}(m,\sigma)$ and $\langle c^\dagger_{k\alpha}c_{k'\alpha'}\rangle=\delta_{\alpha\alpha'}\delta(k-k')n_F(\epsilon_k^{(\alpha)})$, we obtain
\begin{align}
    \nonumber
    \langle a^\dagger_{m\sigma}a_{m'\sigma'}\rangle &= \sum_\alpha \int_{-\pi}^\pi dk\\
    &\hspace{-1.5cm}
     \times\psi_k^{(\alpha)}(m,\sigma)^* \psi_k^{(\alpha)}(m',\sigma') n_F(\epsilon_k^{(\alpha)})\,, 
\end{align}
which, for the case of small temperatures $T\ll\Delta$ and by inserting (\ref{eq:Bloch_eigenstates}), can be written as
\begin{align}
    \nonumber
    \langle a^\dagger_{m\sigma}a_{m'\sigma'}\rangle &= \sum_{\alpha=1}^\nu \int_{-\pi}^\pi \frac{dk}{2\pi}\\
    \label{eq:propagator}
    &\hspace{-1.5cm}
     \times\chi_k^{(\alpha)}(j,\sigma)^* \chi_k^{(\alpha)}(j',\sigma') e^{-ik(n-n')}\,. 
\end{align}
Here $\sum_{\alpha=1}^\nu$ denotes the sum over the occupied bands. The Bloch vectors $\chi_k^{(\alpha)}$ depend parametrically on $e^{ik}$ and on the dispersion $\epsilon_k^{(\alpha)}$. The latter has a branching point in the complex plane with an imaginary part $\text{Im}(k)\sim a/\xi_c$ \cite{rehr_kohn_prb_74,kallin_halperin_prb_84,he_vanderbilt_prl_01,pletyukhov_etal_prb_20}, where $\xi_c$ is some length scale averaged over all channel indizes. Closing the integration contour in the complex plane this gives the exponential decay of the propagator $\sim e^{-Z|n-n'|a/\xi_c}$, leading via (\ref{eq:C_U=0_clean}) to the corresponding exponential decay (\ref{eq:C_exp}) of the correlation function. For scales $|n-n'|a\sim \xi_c$ and $\xi_c\gg a$ we get $k\sim |n-n'|^{-1}\sim a/\xi_c\ll 1$ such that all $e^{ik}\approx 1$ and $\epsilon_k^{(\alpha)}\sim \Delta$. Therefore, the integral $\int dk \sim a/\xi_c$ and the propagator scales in the same way. For very short scales $|n-n'|\sim O(1)$ and $\xi_c\gg a$, the gap is not relevant and the propagator is of $O(1)$. The same occurs if $|n-n'|\sim O(1)$ and $\xi_c\sim a$. This proves (\ref{eq:C_scale}).

Finally we note that it is quite useful to express the propagator (\ref{eq:propagator}) and the fluctuations via the density matrix introduced in Ref.~[\onlinecite{he_vanderbilt_prl_01}]. For the propagator we get
\begin{align}
    \label{eq:prop_dm}
    \langle a^\dagger_{m\sigma} a_{m'\sigma'} \rangle = \frac{Z}{2 \pi} \int_{-\pi/Z}^{\pi/Z} d \bar{k}\,\,
    \hat{n}_{j'\sigma',j\sigma} (\bar{k})  \, \,e^{- i \bar{k} (m-m')},
\end{align}
where $\bar{k}=k/Z$,
\begin{align}
    \label{eq:bar_chi}
    \bar{\chi}_{\bar{k}}^{(\alpha)} (j,\sigma) = e^{i \frac{k}{Z} (Z-j)} \chi_{k}^{(\alpha)}(j,\sigma) \,,
\end{align}
and the density matrix (written as an operator in unit cell space)
\begin{align}
    \label{eq:dm_zero_T}
    \hat{n} (\bar{k}) = \sum_{\alpha=1}^{\nu} | \bar{\chi}_{\bar{k}}^{(\alpha)}\rangle \langle \bar{\chi}_{\bar{k}}^{(\alpha)}|
    = \hat{n}(\bar{k})^\dagger\,.
\end{align}
We note the properties
\begin{align}
    \hat{n}(\bar{k})^2 &= \hat{n}(\bar{k})\,, \label{eq:proj_prop} \\
    \text{tr} \, \hat{n}_(\bar{k}) &= \nu \,, \label{eq:tr_prop}
\end{align}
which follow immediately from the orthogonality relation $\langle\bar{\chi}_{\bar{k}}^{(\alpha)}|\bar{\chi}_{\bar{k}}^{(\alpha')}\rangle = \delta_{\alpha \alpha'}$.

Using (\ref{eq:prop_dm}) one finds after integration by parts
\begin{align}
    \nonumber
    & (m-m') \langle a^\dagger_{m\sigma} a_{m'\sigma'} \rangle = -i \frac{Z}{2 \pi} \int_{-\pi /Z}^{\pi /Z} d \bar{k} \\
    &\hspace{1cm}
    \times  \Big[\partial_{\bar{k}} \hat{n}_{j'\sigma',j\sigma} (\bar{k}) \Big] e^{- i \bar{k} (m-m')}\,.
    \label{eq:amm_zeroT}
\end{align}
Therefore, from (\ref{eq:QB_fluc_BBC}) and (\ref{eq:C_U=0_clean}) one can write the fluctuations in terms of the density matrix as
\begin{align}
    \label{eq:QB_DM} 
    l_p \, \Delta Q_B^2 =  \frac{a}{2}   \int_{-\pi /Z}^{\pi /Z} \frac{d \bar{k}}{2 \pi}  \text{tr} \{ [ \partial_{\bar{k}} \hat{n} (\bar{k}) ]^2 \} \,.
\end{align}

The formula \eqref{eq:QB_DM} can be equivalently rewritten in terms of the Bloch momentum $k$ (instead of $\bar{k}$):
\begin{align}
    \label{eq:QB_DM_k}
    l_p \, \Delta Q_B^2 & =  \frac{Za}{2}   \int_{-\pi }^{\pi} \frac{d k}{2 \pi}  \text{tr} \{ [ \partial_{k} \hat{n} (k) ]^2 \}. 
\end{align}

At finite temperature we have to use the following form of the density matrix  
\begin{align}
    \label{eq:dm_finite_T}
    \hat{n}(\bar{k}) \rightarrow \tilde{n}(\bar{k}) = \sum_{\alpha} | \bar{\chi}_{\bar{k}}^{(\alpha)}\rangle \langle \bar{\chi}_{\bar{k}}^{(\alpha)}|
    \,n_F (\epsilon_{\bar{k}}^{(\alpha)})\,.
\end{align}

\subsection{Example: Rice-Mele model}
\label{sec:rice_mele}

The Rice-Mele model is defined by $Z=2$ with two hoppings $t_{1/2}>0$ and staggered on-site potentials $v=v_1=-v_2$.
The Bloch Hamiltonian $h_k$ in the two-dimensional unit cell reads 
\begin{align}
    h_k = \left(\begin{array}{cc} v & - t_1 -t_2 e^{-i k} \\ - t_1 -t_2 e^{i k} & - v \end{array} \right)
\end{align}
and has eigenvalues
\begin{align}
    \epsilon_k^{(\pm)} = \pm \epsilon_k = \pm \sqrt{\Delta^2+ 4 t_1 t_2 \cos^2 \frac{k}{2}},
\end{align}
where $\Delta = \sqrt{v^2 + (t_1 - t_2)^2}$ is half the energy gap $E_g = 2\Delta$ between the valence $\epsilon_k^{(-)}$ and the conduction $\epsilon_k^{(+)}$ bands. The corresponding eigenstates read
\begin{align}
    \chi_k^{(\pm)} = \frac{1}{\sqrt{2 \epsilon_k (\epsilon_k \mp v)}} \left( \begin{array}{c} t_1 + t_2 e^{-i k} \\  v \mp \epsilon_k \end{array}\right).
    \label{eq:eigst_RM}
\end{align}

The expectation value for the boundary charge has been analysed in all detail in Ref.~[\onlinecite{lin_etal_prb_20}]. 
To calculate the fluctuations of the boundary charge we first insert the eigenstates (\ref{eq:eigst_RM})  in (\ref{eq:bar_chi}), and find from \eqref{eq:dm_zero_T} for the density matrix of the valence band
\begin{align}
\hat{n} (k)  &= \frac{1}{2 \epsilon_k } \left( \begin{array}{cc} \epsilon_k - v & t_1 e^{\frac{i}{2} k} + t_2 e^{-\frac{i}{2} k} \\ t_1  e^{-\frac{i}{2} k}+ t_2 e^{\frac{i}{2} k} & \epsilon_k + v \end{array} \right)\,.
\label{eq:n_RM}
\end{align}
Computing the $k$-derivative we obtain
\begin{align}
 & \partial_k \hat{n} (k) \label{eq:n_der_RM} \\
 =&  \partial_k (\frac{1}{2 \epsilon_k }) \left( \begin{array}{cc} - v & t_1 e^{\frac{i}{2} k} + t_2 e^{-\frac{i}{2} k} \\ t_1 e^{-\frac{i}{2} k} + t_2 e^{\frac{i}{2} k} & v  \end{array} \right) \nonumber \\
 +&  \frac{i}{4 \epsilon_k } \left( \begin{array}{cc} 0 &  t_1 e^{\frac{i}{2} k}- t_2 e^{-\frac{i}{2} k} \\ -  t_1 e^{-\frac{i}{2} k} + t_2  e^{\frac{i}{2} k} & 0 \end{array} \right).\nonumber
\end{align}
Using $ \partial_k (\frac{1}{ \epsilon_k }) = \frac{t_1 t_2 \sin k}{\epsilon_k^3}$, we evaluate
\begin{align}
 & \frac12 \int_{-\pi}^{\pi} \frac{d k}{2 \pi} \text{tr} \{ [\partial_k \hat{n} (k) ]^2 \} =  \int_{-\pi}^{\pi} \frac{d k}{2 \pi} \frac{1}{16 \epsilon_k^6}\nonumber \\ 
 & \times [ (\Delta^2 + 4 t_1 t_2)^2 (t_1-t_2)^2 \cos^2 \frac{k}{2} \nonumber \\
 &+ \Delta^4 (t_1 + t_2)^2 \sin^2 \frac{k}{2} + 4 v ^2 t_1^2 t_2^2  \sin^2 k ]   .
\label{eq:mqbZ2app}
\end{align}
Performing this integral, we obtain from (\ref{eq:QB_DM_k})
\begin{align}
l_p \, \Delta Q_{B}^2 &= a\frac{t_1^2 + t_2^2}{8 \Delta} \frac{1}{\sqrt{\Delta^2 + 4 t_1 t_2}} .
\label{eq:mqbZ2}
\end{align}
In the wide-band limit $\frac{t_1+t_2}{2} \equiv t \gg \Delta$ we estimate
\begin{align}
\label{eq:mqbZ2_WB}
l_p \, \Delta Q_{B}^2  \approx \frac{ta}{8 \Delta} = \frac{v_F}{16 \Delta}\,, 
\end{align}
with $v_F=2ta$.

\subsubsection{Finite temperature}
\label{sec:rice_mele_finite_T}

At finite temperature (but still at fixed particle number) we replace $\hat{n} (k) $ by $\tilde{n} (k)$ defined in \eqref{eq:dm_finite_T} (here we use $k$ instead of $\bar{k}=k/2$). For a two-band model ($\alpha = \pm $) with the particle-hole symmetry $\epsilon_{k}^{(-)}= - \epsilon_{k}^{(+)} \equiv - \epsilon_{k}$, we use the completeness relation $\sum_{\alpha= \pm} | \bar{\chi}_{k}^{(\alpha)} \rangle \langle  \bar{\chi}_{k}^{(\alpha)} |= \hat{1}$ in order to simplify \eqref{eq:dm_finite_T}
\begin{align}
     \tilde{n} (k) &=   n_F (-\epsilon_{k}) \,  \hat{n}(k) +  n_F (\epsilon_{k}) \, (\hat{1} - \hat{n}(k) ) . \label{ntilde_twoband_PHsymm}
\end{align}
Hence
\begin{align}
    \partial_k \tilde{n} (k) = [n_F (- \epsilon_k) - n_F (\epsilon_k)]  \partial_k\hat{n} (k) \nonumber \\
    + \{[ \hat{1}-\hat{n} (k) ] n'_F ( \epsilon_k) - \hat{n} (k) n'_F (- \epsilon_k) \} \frac{d \epsilon_k}{d k} .
\end{align}
Squaring this expression and using the properties \eqref{eq:proj_prop}, 
\eqref{eq:tr_prop} as well as $\text{tr} [\hat{n} (k) \partial_k\hat{n} (k)] =\text{tr} [(\hat{1} -\hat{n} (k)) \partial_k\hat{n} (k)]=0$, we find the wide-band limit expression
\begin{align}
    & l_p \,\Delta Q_{B}^2  \approx \nonumber \\
    &\approx  Za \int_{-\infty}^{\infty} \frac{d k}{2 \pi}\left\{  \frac{v_F^2 \Delta^2}{4 a^2 \epsilon_k^4 } [n_F (- \epsilon_k) -n_F ( \epsilon_k)]^2 \right. \nonumber \\
    & \left. + \frac12 [ (n'_F (- \epsilon_k))^2 + (n'_F ( \epsilon_k))^2] \left( \frac{d \epsilon_k}{d k}\right)^2\right\}.
    \label{eq:fluct_RM_T}
\end{align}
Note that the first term in the integrand follows from its finite-band counterpart in \eqref{eq:mqbZ2app} in the limit under consideration.

Choosing $\mu =0$, we evaluate \eqref{eq:fluct_RM_T}
\begin{align}
    & l_p \, \Delta Q_{B}^2 \approx \nonumber  \\
    &\approx  2\frac{v_F}{\Delta} \int_{0}^{\infty} \frac{d x}{2 \pi}\left\{  \frac{\tanh^2 (\frac{ \Delta}{2 T} \sqrt{x^2 +1}) }{2 (1+x^2)^2}  \right. \nonumber \\
    & \left. + \frac{(\Delta /T)^2}{8 \cosh^4 (\frac{\Delta}{2 T} \sqrt{x^2+1})}  \frac{x^2}{x^2+1} \right\},
    \nonumber
\end{align}
with the rescaled integration variable $x = \frac{v_F k}{a\Delta}$. This function decays monotonically in $T$, and at $T \gtrsim \Delta$ it behaves like $\sim \frac{v_F}{12 \pi T}$. At zero $T$ we recover \eqref{eq:mqbZ2_WB}, finding additionally the low-temperature correction
\begin{align}
    \approx  - \frac{2v_F}{ \Delta \sqrt{2 \pi  \Delta /T}} e^{-  \Delta /T} .
\end{align}

\subsection{Low-energy theory for single channel models}
\label{sec:C_dirac_noninteracting}

For the case of a noninteracting and clean single channel lattice model in the limit of small gap $E_g \ll W$, where $\xi_g\gg a$, one can describe the low-energy physics by an effective Dirac Hamiltonian in $1+1$ dimensions \cite{pletyukhov_etal_prr_20} 
\begin{align}
\nonumber
H_{\rm bulk} &= \int dx \,\underline{\psi}^\dagger(x) \left\{-i v_F \partial_x \sigma_z \right.\\
\label{eq:H_eff}
&\hspace{0cm}
\left.  +\Delta \cos{\gamma}\, \sigma_x - \Delta \sin{\gamma}\, \sigma_y\right\} \underline{\psi}(x)\,,
\end{align}
where $E_g=2\Delta$ is the gap size, $k_F=\pi\nu/(Za)$ is the Fermi momentum at which the gap opens, $v_F=2ta\sin(k_F a)$ denotes the Fermi velocity ($t~\sim W$ is the average hopping), and $\underline{\psi}(x)$ is a two-component field consisting of slowly varying right- and left-moving fields $\psi_\pm(x)$ such that the physical field operator can be expressed as 
\begin{align}
    \label{eq:psi_dirac}
    \psi(x) = \sum_{p=\pm}\psi_p(x)e^{ip k_F x}\,.
\end{align}
The variable $\gamma$ describes the phase of the order parameter such that $\Delta e^{i\gamma}$ describes the transition matrix element from $-k_F$ to $k_F$, see Ref.~\onlinecite{pletyukhov_etal_prr_20}.

The Dirac model has two bands with dispersion $\pm\epsilon_k=\pm\sqrt{v_F^2k^2 + \Delta^2}$. For a chemical potential in the gap and $T\ll\Delta$ all states of the valence band are filled. The eigenstates of the valence band states are given by 
\begin{align}
    \label{eq:eigenstates_dirac}
    \underline{\psi}_k(x) = \frac{1}{\sqrt{2\pi N_k}}
    \begin{pmatrix}-\Delta e^{i\gamma} \\ v_{F}k+\epsilon_{k} \end{pmatrix} e^{ikx} \,,
\end{align}
with normalization factor $N_k=\Delta^2 + (v_F k + \epsilon_k)^2 = 2\epsilon_k (\epsilon_k + v_F k)$. Using this result one can straightforwardly calculate the propagators 
\begin{align}
    \nonumber
    \langle \psi^\dagger_p(x)\psi_{p'}(x')\rangle &= 
    \int dk \langle \psi^\dagger_{k,p}(x)\psi_{k,p'}(x')\rangle \\
    \nonumber &\hspace{-2cm}
    = \frac{1}{2}\delta(x-x')\delta_{pp'} + \frac{1}{4\pi\xi_g} \Big[i\sigma_z K_1\left(\frac{x-x'}{2\xi_g}\right) \\
    \label{eq:Rl_propagators}
    & \hspace{-1.8cm}
    -(\cos{\gamma}\,\sigma_x + \sin{\gamma}\,\sigma_y) K_0\left(\frac{x-x'}{2\xi_g}\right)\Big]_{pp'}\,,
\end{align}
where $K_\nu$ denotes the modified Bessel function of the second kind, $\sigma_i$ are the Pauli matrices, and $\xi_g=\frac{v_F}{2\Delta}=\frac{v_F}{E_g}$ is the exponential decay length. Using (\ref{eq:psi_dirac}) and omitting the divergent contribution at $x=x'$ (which can not be determined from a low-energy model) we find for the density-density correlation function the form
\begin{align}
    \nonumber
    C_{\rm bulk}(x,x') &= - |\langle \psi^\dagger(x)\psi(x')\rangle|^2 \\
    \nonumber
    &\hspace{-1.5cm}
    = - \frac{1}{4\pi^2\xi_g^2}\Big\{K_1\left(\frac{x-x'}{2\xi_g}\right)\sin[k_F(x-x')]  \\
    \label{eq:C_dirac}
    &\hspace{-1cm}
    - K_0\left(\frac{x-x'}{2\xi_g}\right)\cos[k_F(x+x')+\gamma]\Big\}^2\,.
\end{align}
To calculate $C^{\rm bulk}_{jj'}\big(Z(n-n')a\big)$ of the original lattice model, we insert $x=ma$ and $x'=m'a$ into this equation. Using $m=Z(n-1)+j$ and $m'=Z(n'-1)+j'$ together with $k_F=\pi\nu/(Za)$, one finds (except for $n=0$ and $j=j'$) the final result
\begin{align}
    \nonumber
    C^{\rm bulk}_{jj'}(Zn) &= - \frac{1}{4\pi^2\xi_g^2} \\
    \nonumber
    &\hspace{-1cm}
    \times\Big\{K_1\left(\frac{Zn+j-j'}{2\xi_g}\right)\sin[\pi\frac{\nu}{Z}(j-j')]  \\
    \label{eq:C_jj'_dirac}
    &\hspace{-1.5cm}
    - K_0\left(\frac{Zn+j-j'}{2\xi_g}\right)\cos[\pi\frac{\nu}{Z}(j+j')+\gamma]\Big\}^2\,.
\end{align}
Comparing this analytical result in the low-energy limit with exact numerical ones for the original lattice model we find for small gaps a surprisingly perfect agreement even for small values of $n$. Using the asymptotic forms 
\begin{align}
    \label{eq:K0_asymptotics}
    K_0(s)&\rightarrow \begin{cases} \sqrt{\frac{\pi}{2|s|}} \, e^{-|s|} & \text{for}\,|s|\gg 1 \\ 
    -\ln{|s|} & \text{for}\, |s|\ll 1 \end{cases}\\
    \label{eq:K1_asymptotics}
    K_1(s)&\rightarrow \begin{cases} \text{sign}(s)\sqrt{\frac{\pi}{2|s|}}\, e^{-|s|} & \text{for}\,|s|\gg 1 \\ 
    \frac{1}{s} & \text{for}\, |s|\ll 1 \end{cases}\,,
\end{align}
we find that the properties (\ref{eq:C_exp}) and (\ref{eq:C_scale}) are fulfilled. Averaging the correlation function over $j$ and $j'$ according to (\ref{eq:C_n}) we find for $|n|\gg 1$ (where we can neglect $j-j'$ in the argument of the Bessel functions in (\ref{eq:C_jj'_dirac})) the compact form 
\begin{align}
    \nonumber
    \bar{C}_{\rm bulk}(x) &\approx \\
    \label{eq:bar_C_dirac} 
    &\hspace{-1.5cm}
    -\frac{1}{8\pi^2\xi_g^2}\left\{\big[K_0\left(\frac{x}{2\xi_g}\right)\big]^2 + \big[K_1\left(\frac{x}{2\xi_g}\right)\big]^2\right\}\,,
\end{align}
with the following asymptotics for small and large $|x|$
\begin{align}
    \nonumber
    \bar{C}_{\rm bulk}(x)&\rightarrow \\
    \label{eq:bar_C_asymptotics}    
    &\hspace{-1cm}
    -\frac{1}{2\pi}\begin{cases} \frac{1}{2\xi_g |x|}e^{-|x|/\xi_g} & \text{for}\,|x|\gg \xi_g \\ 
    \frac{1}{\pi x^2} & \text{for}\, |x|\ll \xi_g \end{cases}\,.
\end{align}

Inserting the result (\ref{eq:bar_C_dirac}) in (\ref{eq:QB_fluc_BBC_low_energy}) gives the following result for the boundary charge fluctuations close to the phase transition point
\begin{align}
    \label{eq:QB_fluc_dirac}
    l_p \, \Delta Q_B^2 = \frac{\xi_g}{8} = \frac{v_F}{16\Delta}\,.
\end{align}
This result generalizes the result (\ref{eq:mqbZ2_WB}) obtained for the Rice-Mele model to any single channel model.

\section{SSH model with disorder}
\label{sec:disorder}

Here we treat the disordered Su-Schrieffer-Heeger (SSH) model in Born approximation to calculate the gap at moderate disorder strength. The infinite bulk model is defined by the single-particle Hamiltonian
\begin{align}
    \label{eq:SSH_disorder}
    h &= h_0 + V\,,\\
    \label{eq:SSH}
    h_0 &= -\sum_m t_m |m+1\rangle\langle m| + {\rm h.c.}\,,\\
    \label{eq:V}
    V &= -\sum_m w_m |m+1\rangle\langle m| + {\rm h.c.}\,,
\end{align}
where $t_m = t_{m+2}$ are alternating hoppings with $t_{1/2}>0$, and $w_m$ describes random bond disorder taken from a uniform distribution $w_m\in [-d_m/2,d_m/2)$, with $d_m=d_{m+2}>0$ describing the strength of disorder on site $m$. The disorder-averaged progagator can be written in terms of the self-energy as
\begin{align}
    \nonumber
    \bar{G}(E) &= \left(\prod_m \frac{1}{d_m}\int_{-d_m/2}^{d_m/2}dw_m\right) \frac{1}{E-h}\\
    \label{eq:G_averaged}
    &=\frac{1}{E-h_0-\Sigma(E)}\,.
\end{align}
From the definition we get the useful properties (for any $E$ in the complex plane)
\begin{align}
    \label{eq:adjoint}
    \bar{G}(E)^\dagger = \bar{G}(E^*) \quad,\quad
    \Sigma(E)^\dagger = \Sigma(E^*)\,,
\end{align}
or 
\begin{align}
    \label{eq:adjoint_2}
    \bar{G}(E)^T = \bar{G}^*(E) \quad,\quad
    \Sigma(E)^T = \Sigma^*(E)\,,
\end{align}
where $A(E)^T$ is the transposed matrix and $A^*(E)$ denotes the conjugate complex of the matrix (without taking the conjugate complex of $E$). Due to translational invariance we can write the SSH Hamiltonian $h_0$ and the self-energy $\Sigma(E)$ in diagonal form with respect to the quasimomentum $k$
\begin{align}
    \label{eq:h_k}
    h_0 &= \int_{-\pi}^\pi dk |k\rangle\langle k| \otimes h_k^{(0)}\,,\\
    \label{eq:sigma_k}
    \Sigma(E) &= \int_{-\pi}^\pi dk |k\rangle\langle k| \otimes \Sigma_k(E)\,,
\end{align}
where 
\begin{align}
    \label{eq:plane_waves}
    \langle n|k\rangle = \frac{1}{\sqrt{2\pi}}e^{ikn} 
\end{align}
are plane waves with respect to the unit cell index $n$, and $h_k^{(0)}$ and $\Sigma_k(E)$ are $2\times 2$-matrices within the $2$-dimensional unit cell space. For the SSH model we get the form 
\begin{align}
    h_k^{(0)} = \left(\begin{array}{cc} 0 & - t_1 -t_2 e^{-i k} \\ - t_1 -t_2 e^{i k} & 0 \end{array} \right)\,.
\end{align}
In this notation matrix elements of the free propagator $g(E)=1/(E-h_0)$ can be written as
\begin{align}
    \label{eq:free_propagator}
    g(E)_{mm'} = \hat{g}(E,n-n')_{jj'}\,,
\end{align}
where $m=2(n-1)+j$, $m'=2(n'-1)+j'$ and
\begin{align}
    \nonumber
    \hat{g}(E,n) &= \int_{-\pi}^\pi \frac{dk}{2\pi}\,e^{ikn}\,\frac{1}{E-h_k^{(0)}}\\
    \label{eq:g(E,n)}
    &\hspace{-1cm}
    = \int_{-\pi}^\pi \frac{dk}{2\pi} \,e^{ikn} \,\frac{E-(t_1+t_2\cos{k})\sigma_x - (t_2 \sin{k}) \sigma_y}
    {E^2 - t_1^2 - t_2^2 - 2t_1 t_2 \cos{k}}
\end{align}
is a $2\times 2$-matrix with $\sigma_{x,y,z}$ denoting the Pauli matrices. 

In standard Born approximation (see, e.g., Ref.~\onlinecite{groth_etal_prl_09}) we find for the nonvanishing matrix elements
\begin{align}
    \label{eq:sigma_born_nondiagonal_1}
    \Sigma(E)_{m,m+1} &= \frac{d_m^2}{12}\,g(E)_{m+1,m}\,,\\ 
    \label{eq:sigma_born_nondiagonal_2}
    \Sigma(E)_{m+1,m} &= \frac{d_m^2}{12}\,g(E)_{m,m+1}\,,\\ 
    \label{eq:sigma_born_diagonal}
    \Sigma(E)_{m,m} &= \frac{d_m^2}{12}\,g(E)_{m+1,m+1} + \frac{d_{m-1}^2}{12}\,g(E)_{m-1,m-1}\,.
\end{align} 
Since, according to (\ref{eq:free_propagator}) and (\ref{eq:g(E,n)}), the diagonal elements $g(E)_{mm}$ of the propagator are independent of $m$, (\ref{eq:sigma_born_diagonal}) describes only a constant shift of the energy leading to the renormalized energy
\begin{align}
    \label{eq:E_renormalization}
    \tilde{E} = E \left(1 - \frac{d_1^2+d_2^2}{12}\int_{-\pi}^\pi 
    \frac{dk}{2\pi} \frac{1}{E^2 - t_1^2 - t_2^2 - 2t_1 t_2 \cos{k}}\right)\,,
\end{align}
such that the averaged propagator in Born approximation can be written as
\begin{align}
    \label{eq:G_born}
    \bar{G}(E) = \frac{1}{\tilde{E}-h(E)}\,,
\end{align}
with $h(E) = h_0 + \Sigma(E)$. Using the matrix elements of the self-energy in Born approximation we can write
\begin{align}
    \label{eq:h(E)}
     h(E) = -\sum_m \Big\{t_m(E) |m+1\rangle\langle m| + t^*_m(E) |m\rangle\langle m+1| \Big\}\,, 
\end{align}
where $t_m(E)=t_{m+2}(E)$ and
\begin{align}
    \label{eq:t1(E)}
    t_1(E) = t_1 - \Sigma(E)_{21} = t_1 - \frac{d_1^2}{12}\,g(E)_{12}\,,\\
    \label{eq:t2(E)}
    t_2(E) = t_2 - \Sigma(E)_{32} = t_2 - \frac{d_2^2}{12}\,g(E)_{23}\,,
\end{align}
and $t^*_j(E)$ follows from the conjugate complex (but leaving $E$ invariant). Using (\ref{eq:free_propagator}) we get $g(E)_{12}=\hat{g}(E,0)_{12}$ and $g(E)_{23}=\hat{g}(E,-1)_{21}$ which, together with (\ref{eq:g(E,n)}), leads to 
\begin{align}
    \label{eq:g_12}
    g(E)_{12} &= \int_{-\pi}^\pi \frac{dk}{2\pi} 
    \frac{t_1 + t_2 e^{ik}}{|t_1+t_2 e^{ik}|^2 - E^2}\,,\\
    \label{eq:g_23}
    g(E)_{23} &= \int_{-\pi}^\pi \frac{dk}{2\pi} 
    \frac{t_2 + t_1 e^{ik}}{|t_2+t_1 e^{ik}|^2 - E^2}\,.
\end{align}
At zero energy we get 
\begin{align}
    \label{eq:g_12_E=0}
    g(0)_{12} &= \frac{1}{t_1}\theta(t_1-t_2)\,,\\ 
    \label{eq:g_23_E=0}
    g(0)_{23} &= \frac{1}{t_2}\theta(t_2-t_1)\,, 
\end{align}
leading to the following final result for the gap in the presence of disorder
\begin{align}
    \nonumber
    E_g &= 2|t_1(0)-t_2(0)| \\
    \label{eq:gap_disorder}
    &= |2(t_1-t_2) - \frac{d_1^2}{6\,t_1}\theta(t_1-t_2) + \frac{d_2^2}{6\,t_2}\theta(t_2-t_1)|\,.
\end{align}
For $t_1>t_2$ this leads to the gap closing condition
\begin{align}
    \label{eq:gap_closing}
    r = \frac{t_1}{t_2}=\frac{1}{2}\left\{1+\sqrt{1+\frac{d_1^2}{3t_2^2}}\right\}\,,
\end{align}
which agrees rather well with the numerical result up to $d_1\sim 2-3$. For $t_1<t_2$, Born approximation turns out to be insufficient, which will be studied in a future work \cite{weber_etal_future}.

\section{Higher dimensions}
\label{sec:quantum_hall_insulator}

In this section we calculate the fluctuations for noninteracting and clean models in higher dimensions $D=2,3$. We start with $D=2$ and combine a standard 2D quantum Hall insulator \cite{thouless_etal_prl_82} with an additional modulation of the on-site potentials and nearest-neighbor hoppings in $x$-direction. Such models were studied extensively in  Refs.~[\onlinecite{lin_etal_prb_20,pletyukhov_etal_prr_20,park_etal_prb_16,thakurathi_etal_prb_18,pletyukhov_etal_prbr_20,pletyukhov_etal_prb_20}] to study the expectation value of the boundary charge. We start from a 2D tight-binding model with sites labeled by $(m,\sigma)$ with $m=1,\dots,N_s$ in $x$-direction and $\sigma=1,\dots,N_\perp$ in $y$-direction. The lattice spacing $a=a_x=a_y$ is assumed to be the same in both directions. We take open boundary conditions in $x$-direction and periodic ones in $y$-direction. We consider a constant magnetic field $B$ perpendicular to the sample in Landau gauge $\underline{A}(m,\sigma)=(0,Bm,0)$. As a result, the hopping in $y$-direction from $(m,\sigma)\rightarrow (m,\sigma+s)$ aquires a phase factor $e^{i2\pi s m/Z}$, where $\lambda_B=Za$ is the magnetic length defined by $\lambda_B/a = Z = \Phi_0/(Ba^2)$, where $\Phi_0=hc/e$ denotes the flux quantum. For simplicity we assume that $Z$ is an integer. If an additional flux $\Phi$ is applied through the hole of the cylinder when the system is deformed to a ring in $y$-direction, an additional phase factor $e^{-is\theta/N_\perp}$ with $\theta=2\pi \Phi/\Phi_0$ has to be considered for the hopping in $y$-direction by $s$ sites. 

Taking real and negative nearest-neighbor hoppings $-t_m=-t_{m+Z}$ in $x$-direction and $-t_y$ in $y$-direction, together with real on-site potentials $v_m=v_{m+Z}$, we arrive at the following 2D tight-binding model
\begin{align}
    \nonumber
    H &= \sum_{m=1}^{N_s}\sum_{\sigma=1}^{N_\perp} v_m a^\dagger_{m\sigma}a_{m\sigma} \\
    \nonumber
    & - \sum_{m=1}^{N_s-1}\sum_{\sigma=1}^{N_\perp} t_m (a^\dagger_{m+1,\sigma}a_{m\sigma}+\text{h.c.})\\
    \label{eq:2D_tight_binding} 
    & - t_y \sum_{m=1}^{N_s}\sum_{\sigma=1}^{N_\perp}
    \Big(e^{i(2\pi m/Z - \theta/N_\perp)} a^\dagger_{m,\sigma+1}a_{m\sigma}  + \,\text{h.c.}\Big)\,,
\end{align}
with $a_{m\sigma}=a_{m,\sigma+N_\perp}$ due to periodic boundary conditions in $y$-direction. Using Fourier transform for the modes in this direction
\begin{align}
    \label{eq:Fourier}
    a_{m\sigma} = \frac{1}{\sqrt{N_\perp}} \sum_{k_y} e^{i k_y a\sigma} c_{m k_y}\,,
\end{align}
with 
\begin{align}
    \label{eq:ky}
    k_y = \frac{2\pi}{N_\perp a} s \quad,\quad s=1,\dots,N_\perp \,,
\end{align}
we can write the Hamiltonian as an independent sum over all transverse quasimomenta
\begin{align}
    \label{eq:H_sum_ky}
    H &= \sum_{k_y} H(k_y) \\
    \nonumber
    H(k_y) &= \sum_{m=1}^{N_s} \bar{v}_m(k_y a +\theta/N_\perp) c^\dagger_{m k_y}c_{m k_y} \\
    \label{eq:H_phi}
    &-\sum_{m=1}^{N_s-1} t_m (c^\dagger_{m+1,k_y}c_{m k_y} + \text{h.c.}) \,,
\end{align}
with
\begin{align}
    \label{eq:v_m_phi}
    \bar{v}_m (\varphi) = v_m - 2 t_y \cos{\left(\frac{2\pi}{Z}m - \varphi\right)}\,.
\end{align}

The charge operator summed over all transverse modes is given by
\begin{align}
    \label{eq:rho_m}
    \hat{\rho}_m = \sum_{\sigma=1}^{N_\perp} a_{m\sigma}^\dagger a_{m\sigma}
    = \sum_{k_y} c_{m k_y}^\dagger c_{m k_y}\,.
\end{align}
Since the Hamiltonian does not couple the transverse modes, the fluctuations of the boundary charge operator (\ref{eq:QB_operator}) can be written as an independent sum over the transverse modes
\begin{align}
    \label{eq:QB_fluc_sum}
    \Delta Q_B^2 = \sum_{k_y} \Delta Q_B^2(k_y)\,,
\end{align}
where $\Delta Q_B^2(k_y)$ describe the boundary charge fluctuations of the effectively $1$-dimensional Hamiltonian $H(k_y)$ at fixed $k_y$. Denoting by $2\Delta(k_y)$ the gap of this Hamiltonian and by $v_F(k_y)$ the Fermi velocity, we can take in the low-energy limit of small gap compared to the band width the result (\ref{eq:QB_fluc_dirac}), leading for $N_\perp\rightarrow \infty$ to the integral
\begin{align}
    \nonumber
    l_p\Delta Q_B^2 &= \sum_{k_y} \frac{v_F(k_y)}{16\Delta(k_y)}\\
    \label{eq:QB_fluc_total}
    &\rightarrow \frac{N_\perp a}{2\pi}\int_{-\pi/a}^{\pi/a} dk_y \frac{v_F(k_y)}{16\Delta(k_y)}\,.
\end{align}

To evaluate the result (\ref{eq:QB_fluc_total}) explicitly, we consider an illustrative example in terms of the Rice-Mele model, with $Z=2$ and $v_m=0$. Using (\ref{eq:v_m_phi}) we obtain
\begin{align}
    \bar{v}_m(\varphi) = (-1)^m v(\varphi)\,\,,\,\, v(\varphi) = - 2t_y \cos(\varphi)\,.
\end{align}
The gap opens up at $k_F=\pi/(2a)$ which corresponds to half-filling. In the low-energy limit $t_y,|t_1-t_2|\ll t=(t_1+t_2)/2$, we obtain $v_F(k_y)=2ta$ and 
\begin{align}
    \label{eq:delta}
    \Delta(k_y) = \sqrt{\big[v(k_y a + \theta/N_\perp)\big]^2 + |t_1-t_2|^2}\,.
\end{align}
Inserting in (\ref{eq:QB_fluc_total}) and calculating the integral leads to the final result
\begin{align}
    \label{eq:QB_fluc_2D}
    l_p\Delta Q_B^2 = \frac{N_\perp}{8\pi}\frac{v_F}{\sqrt{4t_y^2 + \Delta^2}} K\left(\frac{4t_y^2}{4t_y^2 + \Delta^2}\right)\,,
\end{align}
where $K(p)=\int_0^{\pi/2}dx \big[1-p\sin^2(x)\big]^{-1/2}$ is the elliptic integral of first kind, and 
\begin{align}
    \label{eq:gap_2D}
    E_g = 2\Delta = 2\ \text{min}_{k_y} \Delta(k_y) = 2|t_1-t_2|
\end{align}
denotes the gap for the $2$-dimensional system. As expected the fluctuations are independent of the phase $\theta$.

For $\Delta\ll t_y$, we can use the asymptotics $K(p)\rightarrow \frac{1}{2}\ln\big(16/(1-p)\big)$ for $p\rightarrow 1$, and obtain a logarithmic scaling of the fluctuations as function of the gap close to the phase transition
\begin{align}
    \label{eq:QB_fluc_zero_delta}
    l_p \, \Delta Q_B^2 \xrightarrow{\Delta\ll t_y} \frac{N_\perp}{16\pi}\frac{v_F}{t_y}\ln\frac{8t_y}{\Delta}\,.
\end{align}

For the Rice-Mele model, we can also use the exact result (\ref{eq:mqbZ2}) to calculate the fluctuations at fixed $k_y$. Performing the integral over $k_y$ this leads to the following result for the total fluctuations
\begin{align}
    \label{eq:QB_fluc_2D_exact}
    l_p \, \Delta Q_B^2 = N_\perp a \frac{4t^2 + \Delta^2}{32\pi t_y^2} I(\frac{\Delta}{2t_y},\frac{t}{t_y})\,,
\end{align}
with 
\begin{align}
    \label{eq:integral}
    I(a,b) = \frac{1}{b\sqrt{1+a^2}} K\left(\frac{b^2-a^2}{b^2(1+a^2)}\right)\,.
\end{align}
To compare with (\ref{eq:QB_fluc_2D}) we can rewrite this result as
\begin{align}
    \label{eq:QB_fluc_2D_exact_explicit}
    l_p\Delta Q_B^2 = \frac{N_\perp}{8\pi}\frac{2ta(1+\frac{\Delta^2}{4t^2})}{\sqrt{4t_y^2 + \Delta^2}} 
    K\left(\frac{t_1 t_2}{t^2}\frac{4t_y^2}{4t_y^2 + \Delta^2}\right)\,,
\end{align}
and find that they agree for $\Delta=|t_1-t_2|\ll t=(t_1+t_2)/2$.

In the low-energy regime one can also analyse the boundary charge analytically. It is given as an independent sum over the boundary charges of the effective one-dimensional models 
\begin{align}
    Q_B(\theta) = \sum_{k_y} Q_B^{\rm 1D}(k_y a + \frac{\theta}{N_\perp})\,.
\end{align}
Using (\ref{eq:ky}) and the periodicity $Q_B^{\rm 1D}(\varphi)=Q_B^{\rm 1D}(\varphi + 2\pi)$, this can also be written as
\begin{align}
    \label{eq:QB_theta}
    Q_B(\theta)= \frac{N_\perp}{2\pi}\sum_{l=-\infty}^\infty e^{il\theta} \int_0^{2\pi}d\varphi \,Q_B^{\rm 1D}(\varphi) e^{-il N_\perp\varphi}\,.
\end{align}
At zero chemical potential the boundary charge of the one-dimensional model can be calculated from the low-energy analysis of Ref.~[\onlinecite{pletyukhov_etal_prb_20}] as
\begin{align}
    Q_B^{\rm 1D}(\varphi) = \frac{\gamma(\varphi)}{2\pi} + \frac{1}{4} - \theta_{\pi/2 < \gamma(\varphi) < \pi} \,,
\end{align}
where $-\pi < \gamma(\varphi) < \pi$ and 
\begin{align}
    \Delta(\varphi) e^{i\gamma(\varphi)} = v(\varphi) + i(t_2-t_1)\,.
\end{align}
Since $t_{1,2}$ are independent of the phase, $\gamma(\varphi)=\gamma(\varphi+2\pi)$ is bounded to an interval smaller than $2\pi$ and the phase factor $e^{i\gamma(\varphi)}$ can not wind around the origin in the complex plane. As a consequence, $Q_B^{\rm 1D}(\varphi)$ is a smooth and periodic function of $\varphi$ and the integral in (\ref{eq:QB_theta}) can only contribute for $l=0$. Therefore, for this special model, the boundary charge is independent of $\theta$ and the Hall current is zero. This result holds not only in the low-energy regime but also at large gap. 

Although the topology does not change at the phase transition point $r=t_1/t_2=1$, there is a discontinuous change of the appearance of edge states in the gap. For $r>1$, there is no edge state for any $k_y$, whereas, for $r<1$, an edge state appears for all $k_y$ at energy $-v(k_y a + \theta/N_\perp) = 2 t_y \cos(k_y a + \theta/N_\perp)$ which never touches the band edges. Therefore, the gap has to close at the transition, and the fluctuations show the same characteristic scaling as for topological phase transitions where the gap closing is induced by a change of the Chern number. 

The above analysis can easily be generalized to a $3D$-system by choosing a nearest-neighbor hopping $t_z$ in $z$-direction and adding a magnetic field of size $B$ in $y$-direction. Omitting the additional flux $\Phi$, we can write the Hamiltonian as an independent sum over all $k_y$ and $k_z$
\begin{align}
    \label{eq:H_sum_ky_kz}
    H &= \sum_{k_y k_z} H(k_y,k_z) \\
    \nonumber
    H(k_y,k_z) &= \sum_{m=1}^{N_s} \bar{v}_m(k_y,k_z) c^\dagger_{m k_y k_z}c_{m k_y k_z} \\
    \label{eq:H_phi_3D}
    &-\sum_{m=1}^{N_s-1} t_m (c^\dagger_{m+1,k_y k_z}c_{m k_y k_z} + \text{h.c.}) \,,
\end{align}
with 
\begin{align}
    \nonumber
    \bar{v}_m(k_y,k_z) = v_m &- 2 t_y \cos{\left(\frac{2\pi}{Z}m - k_y a\right)}\\
    \label{eq:bar_v_3D}
    &- 2 t_z \cos{\left(\frac{2\pi}{Z}m - k_z a\right)}\,.
\end{align}
Taking again the special case $Z=2$ and $v_m=0$, we obtain for each fixed $(k_y,k_z)$ an effective Rice-Mele model with potential $v(k_y,k_z)=\bar{v}_1(k_y,k_z)=-\bar{v}_2(k_y,k_z)$, where
\begin{align}
    \label{eq:RM_3D}
    v(k_y,k_z) = 2t_y \cos(k_y a) + 2t_z \cos(k_z a)\,. 
\end{align}
Using the exact result (\ref{eq:mqbZ2}) for the calculation of the fluctuations at fixed $(k_y,k_z)$ and integrating over the transverse momenta one arrives straightforwardly at the following result for the total fluctuations
\begin{align}
    \label{eq:QB_fluc_3D_exact}
    l_p \, \Delta Q_B^2 = N_\perp^2 a^2 \frac{4t^2 + \Delta^2}{32\pi^2 t_y^2} \int_0^{\pi/a} dk_z I\left(g(k_z),h(k_z)\right)\,,
\end{align}
with
\begin{align}
    \label{eq:g}
    g(k_z) &= \frac{1}{2t_y}\sqrt{t_z^2 \cos^2(k_z a) + \Delta^2}\,,\\
    \label{eq:h}
    h(k_z) &= \frac{1}{2t_y}\sqrt{t_z^2 \cos^2(k_z a) + 4 t^2}\,,
\end{align}
and $I(a,b)$ is defined in (\ref{eq:integral}). This leads to a convergent result for the fluctuations in the zero gap limit $\Delta\rightarrow 0$. 

\begin{figure}[t]
    \centering
	 \includegraphics[width =\columnwidth]{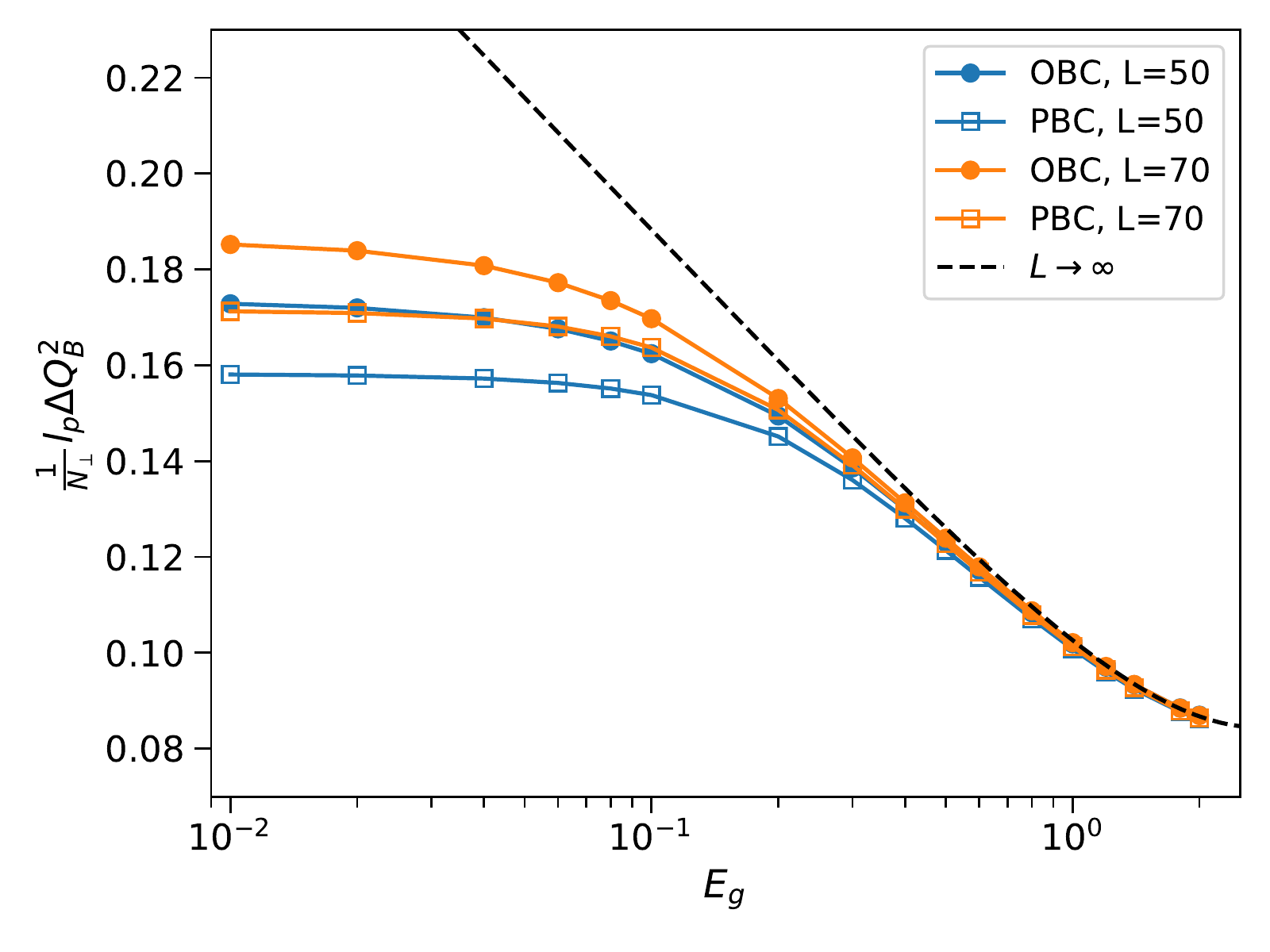}
	  \caption{{\color{black} Comparison of different system sizes  $L=N_\perp a=N_s a$ and boundary conditions of the two dimensional model discussed in the text. The parameters are given by $Z=2$, $v_m=0$, $t_{1/2}=1.0\pm \Delta/2$, $t_y=1.0$, $\theta=0$, $L_p=L/2$ and $l_p=(24, 34)$ whereby $\Delta$ is varied in the interval $[0.005, 1.0]$.}
	  }
	  \label{fig:2DBoundary} 
\end{figure}

{\color{black} To access the influence of different boundary conditions at finite system size we consider the two-dimensional model described above on a square lattice with $L=N_\perp a=N_s a$. Figure \ref{fig:2DBoundary} summarizes the finite size scaling (symbols) compared to the infinite limit (dashed line) for open (OPC) and periodic (PBC) boundary conditions. As the system size is increased the universal result is approached for increasingly smaller gaps $E_g$.

}

\section{Special phase transitions}
\label{sec:special}

Sometimes an expansion of the Bloch Hamiltonian $h_k$ around the quasimomentum where the gap opens does not contain linear terms due to special symmetry conditions. An exemplary model of this type has been discussed in Ref.~[\onlinecite{hine_foulkes_jpc_07}]. It contains only constant and quadratic terms in $k$:
\begin{align}
    h_k = \left( \begin{array}{cc} \Delta + \alpha k^2 & \Gamma k^2 \\ \Gamma k^2 & - \Delta -\beta k^2 \end{array} \right).
    \label{quadr_gap}
\end{align}
The gap between the two bands is $E_g=2 \Delta$. The Bloch eigenstates depend only on the combination $\frac{\Delta}{k^2}$. It follows immediately that the imaginary part $\kappa$ of the branching point appears to be $\kappa \sim \sqrt{E_g}$. Thus, $l_p\Delta Q_B^2\sim E_g^{-1/2}$. 

This can be generalized to the case where the minimal nonvanishing order of $k$ in the expansion of $h_k$ is $l$. This gives straightforwardly the scaling $l_p\Delta Q_B^2\sim\xi \sim E_g^{-1/l}$.


We note in passing that in quasi-two-dimensional models the band structure of the form \eqref{quadr_gap} can be realized in the multilayer graphene with a special stacking \cite{min_macdonald_ptp_08}. However, as has been shown in Section~\ref{sec:quantum_hall_insulator}, for a $2D$-system the fluctuations follow from an integration over the transverse momentum $k_y$. Assuming close to the gap opening a general dispersion relation for the conduction band of the form
\begin{align}
    \label{eq:dispersion_exotic}
    \epsilon(k_x,k_y) = (\Delta^{2/l} + c k^2)^{l/2}\,,
\end{align}
the solution of $\epsilon(k_x,k_y)=0$ for fixed $k_y$ gives in the complex plane the solution $k_x= 2i/\xi(k_y)$, with 
\begin{align}
    \label{eq:xi_exotic}
    \xi(k_y) = \frac{2\sqrt{c}}{\sqrt{\Delta^{2/l}+ck_y^2}}\,.
\end{align}
The gap at fixed $k_y$ is given by $E_g(k_y)=2\epsilon(0,k_y)=2(\Delta^{2/l} + c k_y^2)^{l/2}$, leading to an exotic scaling for the fluctuations of $Q_B$ at fixed $k_y$
\begin{align}
    \label{eq:QB_fluc_exotic_ky}
    l_p \Delta Q_B^2(k_y) \sim \xi(k_y) \sim E_g(k_y)^{-1/l}\,.
\end{align}
However, this does not change the logarithmic scaling of the total fluctuations in terms of the overall gap $E_g=2\Delta$
\begin{align}
    \label{eq:QB_fluc_exotic_total}
    l_p \Delta Q_B^2 \sim \int_{-\pi/a}^{\pi/a} dk_y \,\xi(k_y) \stackrel{E_g\rightarrow 0}{\sim} |\ln(E_g/W)|\,,
\end{align}
where $W$ defines the high-energy cutoff in terms of the band width.


\bibliography{citations_3}